# A Survey on Security Issues of 5G NR: Perspective of Artificial Dust and Artificial Rain

Misbah Shafi[a], *Member, IEEE*, Rakesh Kumar Jha[b,*], *Senior Member, IEEE*, Manish Sabraj[c]
[a,b,c]Department of Electronics and Communication, Shri Mata Vaishno Devi University, J&K India

*Abstract*—5G NR (New Radio) incorporates concepts of novel technologies such as spectrum sharing, D2D communication, UDN, and massive MIMO. However, providing security and identifying the security threats to these technologies occupies the prime concern. This paper intends to provide an ample survey of security issues and their countermeasures encompassed in the technologies of 5G NR. Further, security concerns of each technology are defined mathematically. Thereby defining the impact on the factors of security. Moreover, a methodology is developed in which the influence on security due to artificially generated rain and artificially generated dust on the wireless communication network is studied. By doing so, an attacking scenario is identified, where a half-duplex attack in D2D communication is attained. Half-duplex attack specifies the attack solely on the downlink to spoof the allocated resources, with reduced miss-rate. Thus, ultra-reliable and adequate advances are required to be addressed to remove the obstacles that create a hindrance in achieving the secured and authenticated communicating network.

## I. Introduction

With the remarkable upsurge of requirements such as user experienced high data rate, low latency, energy efficiency, spectral efficiency, UDN, coverage reliability, mobility in the wireless communication network, 5G emerges as a satisfying approach. In addition to these mobile broadband services, several other services depending upon the usage scenario are predicted to be supported by 5G, including ultra-reliable, gigantic machine type information exchange, boosted mobile broadband, and truncated latency communications [1]. The significant principle of next generation wireless communication network lies in reconnoitering unused mm-wave high frequency band, ranging from 3~300GHz [2]. To justify the ten times more connectivity density of 5G wireless networks than that of 4G, the target connectivity density of 5G is essentially not to be less than $10^6$ / $km^2$ [4]. Therefore, to content the stringent challenges of 5G wireless network communication, advanced research with respect to various characteristic features are required to be observed. Mobile edge computing is considered as one of the up-to-date paradigms to solve the issue of burdensome computations and decrease latency more significantly [3], [5], [6]. Power optimization is another step towards the field of research for improving the representative factor of energy efficiency, such as the green radio project. For the parameter of energy efficiency, it is envisaged that 5G is likely to be 100x times (10 mW/Mbps/sec in IMT-advance) more energy efficient in future IMT than in IMT-2000(100mW/Mbps/sec) [7]. Radio resource management is another challenging field of research associated with the next generation 5G communication network. It includes spectrum allocation, user association, and management of power [8]. Communication efficiency is another aspect that has grabbed serious attention in the field of research. To encounter the communication efficiency in 5G, various techniques such as spectrum using NOMA, network-assisted interference suppression and cancellation, coordinated multipoint joint transmission reception, three-dimensional MIMO, full dimensional MIMO are required to be improvised [9]. In addition to it, security is another challenging feature that occupies immense attention in the field of research.

With the enhancement of incipient technologies and the massive progression in the figure of devices, the number of small cell BSs is increased. These are arranged with the centralized macro BS supporting 1000x capacity, therefore, creating an increased number of handovers. It results in the foundation of feasible sites for the malicious attacker, thus creating a threat to the field of security [10]-[12]. Moreover, next generation wireless communication networks (5G) introduces flat architectures, broad involvement of cloud in network communication, and processing, which enhances the susceptibility of attackers in the network [13]. Also, the network of 5G is considered as a heterogeneous network, including Wi-Fi hotspots, microcells, small cells, femtocells. Though it provides cost efficiency but increases the probability of attacks due to the incremented number of breaches in the architecture of the network [14]. Consequently, advanced solutions are required to be conceived to fulfill the measures of security. The measures of cybersecurity include confidentiality (identity protection, communication confidentiality, and location privacy), authentication (authorized service access), integrity, and availability [15]. Therefore the accuracy of these parameters is required to be fulfilled. For accurate security of the network, the time consumed to detect the existence of malicious attackers in the network must be less than the rate of transmission. This criterion is required to prevent the stark loss that may be caused in the network due to the entry of malicious attackers [16].

From preceding years, research on the security of the next generation 5G wireless communication network has revealed various security issues. Numerous attacks are observed to affect the parameters of security. Various countermeasures, including the physical layer security measures [17]-[19], were proposed to thwart these attacks. Conversely, considering security in a systematized manner is still a void to be filled in the field of research. The technology of IoT involves the number of sensors to satisfy a variety of applications. It acts as an evolving technology in 5G architecture. The communication in IoT takes place via inter mediators (gateway) involving a huge number of devices estimated to be 50 billion by 2020. Consequently, gives rise to the number of attacking breaches by the large traffic created by numerous IP devices. Thus, secure communication mechanisms are required to be developed with low latency, less complexity, and more energy efficiency, improved quality of service, throughput, etc. These parameters achieve proper balance in various security mechanisms. Considering the research efforts and security specification of last year, a systematized approach to project security breaches in different technologies is identified in order to encounter challenges for future particularization. This paper provides a comprehensive and unified survey on the security of 5G technologies. The scope of the survey in different technologies of research is to occupy the blank void between the researched topics of security and to exclude the legacies of insecure WCN.

*A. Contribution:*

In summary, the primary contribution of this paper are mentioned below to provide readers with a better understanding

- We develop a systemized survey of various security mechanisms in different technologies of 5G. Starting from the security perspective in massive MIMO, spectrum sharing, D2D, beamforming, UDN followed by IoT. A comprehensive picture of security research is summarized along with their mathematical security models.
- A system model is proposed to thrive the impact of rain on the security of WCN, involving the scenario of AR to investigate the analysis of the attacking setup.
- The impact of dust on the security of WCN is observed by involving the mechanism of AD. Based on these tractable scenarios, attenuation due to AR and AD are evaluated
- A framework of attack in the form of a half-duplex is proposed. Further, the possibility of such an attack on the basis of the hit rate in comparison to the conventional attack is evaluated.

The organization of the paper is mentioned as: section II defines the security background of 5G technologies along with their mathematical security models. Section III depicts the system model involving the artificial rain, artificial dust, and Half Duplex attack along with mathematical calculations. Section IV signifies the challenges in 5G WCN with a security perspective, which are required to be addressed for secure communication. Section V characterizes the conclusion obtained from the paper, respectively. A list of current security projects in WCN and the list of acronyms are mentioned in the appendix.

## II. SECURITY BACKGROUND OF 5G TECHNOLOGIES

In this section, we describe the security aspect of crucial technologies incorporated in a 5G wireless communication network, as shown in Fig. 1. The impact on security, and identifying generic solutions of security enhancement for each technology. The mathematical aspect of security is expressed for each technology. The 5G wireless communication network consists of the following technologies:

*A. Massive MIMO*

Massive MIMO is most widely studied and considered as the fundamental technology which provides the basis for research in the 5G generation and the next generation WCN technology to satisfy various rising demands of service. Its main advantage involves the potential of high spectral efficiency, energy efficiency, and required service demands. Irrespective of its numerous supporting applications several limitations bound its number of application such as the interference created due to the numeral of antennas existing in the network, accessibility of channel state information, possibility of realistic approach for deployment of the network antennas, pilot contamination, spoofing attack, MIM (Man in the Middle) attack, and DoS attack. These challenges give rise to a number of issues; however, security issues occupy the immense interest due to the plausibility of an eavesdropper to intercept the information. Therefore, security management mechanisms are required to enhance protection from a malicious attacker.

Various approaches were formulated to advance the security of the network in the technology of massive MIMO. It utilizes a spatial modulation scheme where the physical layer encryption technique is followed to increase the security performance of the network along with the enhancement of other parameters such as energy efficiency and reliability [22]. However, for physical layer encryption, the complexity of decoding is examined by mapping massive MIMO to a lattice and therefore providing complexity to both quantum and classical computers [23]. Secure physical layer multicast transmission where the BS is known of the statistical CSI of the valid and invalid user with consideration of lower bound of secrecy rate of a multicast system [24]. Spoofing attacks such as pilot spoofing attacks [25], bandwidth spoofing attacks [12] are prevented by addressing the techniques of PLA. PLA in MIMO is examined on the basis of radio channel information, and interaction between multiple antennas with a receiver is formulated. Consequently, the spoofing attack is detected by using Q-learning [26]. One more approach to detect active spoofing attacks includes the criterion of MDL. The scheme of detection involves the allocation of partial power to place over an arbitrary sequence of detection on the sequence of training of a valid receiver [27].

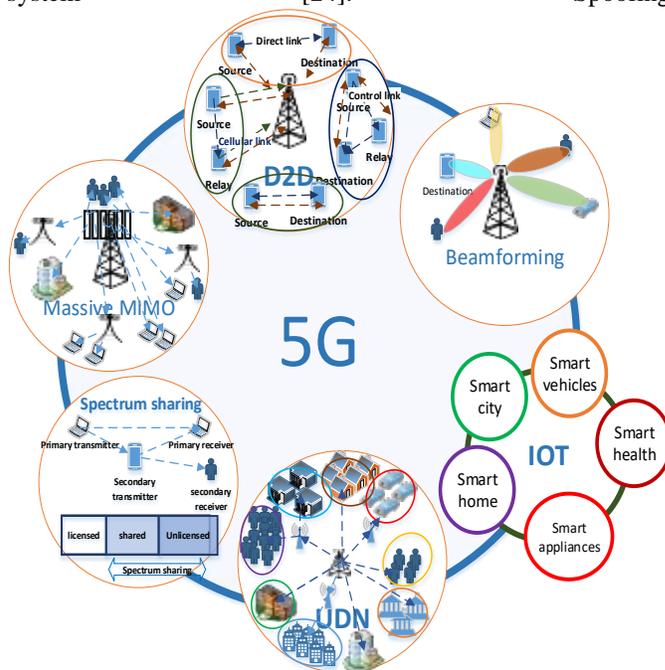

Fig. 1. Generalized 5G architecture

The BS (Base Station) procures the CSI from the uplink transmission of pilots for various receivers. It gives rise to the susceptibility of contamination of the pilot signal based on the estimation of CSI. Therefore, an attack can be initiated, where an attacker targets at reducing the sum rate of downlink transmission by incorporating the contamination of uplink pilots [28]. One recent effort towards the solution of a pilot contamination attack is the uncoordinated mechanism of frequency shift [29]. Another resolution to the security of massive MIMO communication incorporates single-cell downlink Time Division Duplexing (TDD) in the presence of an eavesdropper. Secure degrees of freedom are obtained even though an eavesdropper is present in the network [30]. For the scenario of the wireless network with the consideration of passive eavesdropper only, where secure communication is acquired by employing the massive MIMO relaying technique under the non-availability of channel state information of eavesdropper [31]. In [32]-[34], linear precoding techniques are utilized for the scenario of a passive eavesdropper for the Gaussian MIMO wiretap channel, and these techniques of linear precoding employed in massive MIMO are mentioned in [35]. However, in [36], both active and passive eavesdropper scenarios were taken into concern. One more approach under the consideration of the non-availability of channel state information of eavesdropper, security performance is enhanced by employing the combination of artificial noise and linear precoding techniques [37], [38]. One more approach to ensure PLS in massive MIMO involves the implementation beamforming technique under the consideration of channel estimation errors [39], [40]. Cooperative beamforming provides security by enabling the confidentiality of the information [41], [42].

Moreover, massive MIMO supported C-RAN (Cloud Radio Access Network), enabling the benefits of secrecy along with energy efficiency [43]. A mechanism of cooperative relay network employing in the presence of an eavesdropper whose location is determined on the basis of the Poisson point process incorporating two protocols for relay viz decode-and-forward and amplify-and-forward, therefore ensuring the security of the network [44]. Another PLS scheme involves the use of a pilot under the consideration of a multi-antenna jammer. The detection method for jamming utilizes the likelihood ratio test for generalized coherence blocks [45], [46]. The mechanism of artificial noise is exercised to enhance the security of the network [47]-[50], where an artificial noise is injected in the downlink pilot signal in order to prevent the intruder from getting the correct information about the state of the channel [51]. Other different mechanisms that were followed to achieve secure communication in massive MIMO include beam domain transmission [52], hardware quality scaling procedure [53], beamforming technique with normalized constrained policy of power [54], amendment of AN (Artificial Noise) with low-resolution DACs (Digital to Analog Converters) [55]. AN facilitated design of jamming in the scenario of Racian fading Channel [56], approach of power scaling [57], AN precoding incorporated with RF chains [58], two AN precoding schemes one for the downlink followed by the second AN scheme for the downlink, phase of payload data transmission [59] and distributed reconfigurable MIMO forming cooperative MIMO system [60]. Additional techniques that provide physical layer security in massive MIMO include MF (Matched Filter) precoding combined with AN [61]. TDD integrated with massive MIMO [62], hybrid massive scheme unified with randomization of a signal and the selection of antenna [63]. Furthermore, other solutions to allay pilot contamination in massive MIMO include pilot allocation on the basis of graph coloring [64], scheme of superimposed pilots [65], game theory method followed by optimization scheme [66], [67], approach of approximate location of mobile devices [68], semi-blind scheme for the estimation of the channel in presence of pilot contamination [69], spatial filtering method [70] and methodology of channel sparsity [71]. The inclusive survey of contamination of the pilot signal in massive MIMO is given in [72]. The generalized security model for massive MIMO system can be demonstrated as:

Consider a massive MIMO system entailing of a BS with $i$ number of antennas and a User Equipment (UE) with $j$ number of antennas where the flat fading channel model from BS to UE and from UE to BS can be symbolized as:

$$H_{i,j} = a_j h_{i,j} b_i \qquad (1)$$

or

$$H_{j,i} = a_i h_{j,i} b_j \qquad (2)$$

where $a_j$, $a_i$, $b_j$ and $b_i$ denotes the response of RF front ends of their respective BS and UE during the phase of transmission and reception. $h_{i,j}$ and $h_{j,i}$ symbolizes propagation channel coefficient from BS to UE and form UE to BS, respectively [73]. However, from [74] propagation of the channel from BS to UE can also be defined as:

$$h_{i,j} = \bar{h}_{i,j} + \tilde{h}_{i,j} \qquad (3)$$

Similarly, from UE to BS can also be defined as:

$$h_{j,i} = \bar{h}_{j,i} + \tilde{h}_{j,i} \qquad (4)$$

Equation (4) can also be represented as:

$$h_{i,j} = |\bar{h}_{i,j}|\exp(j2\pi\alpha_{i,j}) + \tilde{h}_{i,j} \qquad (5)$$

where $\bar{h}_{i,j}$ represents the component of the intra array channel model caused by mutual coupling and $\tilde{h}_{i,j}$ models multipath component of the channel apart from mutual coupling, $|\bar{h}_{i,j}|$ and $\alpha_{i,j}$ designates the magnitude and phase component of $\bar{h}_{i,j}$ correspondingly. The received signal can be obtained as:

$$y_j = H_{j,i}x_i + n_j \qquad (6)$$

From equation (2), equation (6) can also be re-written as:

$$y_j = a_i h_{j,i} b_j x_i + n_j \qquad (7)$$

Using equation (4), the above equation can be given as:

$$y_j = a_i b_j (\bar{h}_{j,i} + \tilde{h}_{j,i}) x_i + n_j \quad (8)$$

$$y_j = a_i b_j \bar{h}_{j,i} x_i + a_i b_j \tilde{h}_{j,i} x_i + n_j \quad (9)$$

From [75], interference due to other users is taken into consideration for the purpose of analysis. Therefore, equation (9) can also be deduced as:

$$y_j = a_i b_j \bar{h}_{j,i} x_i + a_i b_j \tilde{h}_{j,i} x_i + n_j + \sum_{k=1}^{j} I_k \quad (10)$$

where $n_j$ represents the AWGN with zero mean and $\sigma^2$ as the noise power, $I$ as the interference due to adjacent users present in the vicinity.

Also, the corresponding total power consumption can be deduced as:

$$O_T = O_{tx} + \sigma^2 + i_p + O_{mc} + O_F \quad (11)$$

where $O_T$ is the total power, $O_{tx}$ is the transmitted power by the BS, $i_p$ is the interference power and is given as $i_p = \sum_{k=1}^{j} |I_k|^2$, $O_{mc}$ power due to mutual coupling of the array, $O_F$ is the front end power and is given by $O_F = |a_i|^2 |b_j|^2$

Now, the SINR between the BS and the authenticated user is given by:

$$SINR = \frac{signal\ power}{noise\ power + mutual\ coupling\ power + interference\ power} \quad (12)$$

$$\varrho_j = \frac{O_{tx} |H_{i,j}|^2}{\sigma_j^2 + i_p + O_{mc}} \quad (13)$$

Similarly, SINR for a malicious intruder is modeled as:

$$\varrho_{in} = \frac{O_{tx} |H_{i,in}|^2}{\sigma_{in}^2 + i_p + O_{mc}} \quad (14)$$

The available SR can be expressed as:

$$SR = (C_j - C_{in})^+ \quad (15)$$

From the theorem of Shannon's capacity

$$C_j = \log_2(1 + \varrho_j) \quad (16)$$

And $\quad C_{in} = \log_2(1 + \varrho_{in}) \quad (17)$

where $C_j$ defines the capacity of the valid user in the absence of an intruder and $C_{in}$ defines the capacity of the malicious intruder. $\varrho_j$ defines the SINR between the BS and authenticated user. $\varrho_{in}$ defines the SINR between the BS and the malicious intruder. $|H_{i,j}|^2$, $|H_{i,in}|^2$ models the magnitude of the channel gain for the valid user and illegitimate user, respectively.

Using values of equation (13) and (14) in equations (16) and (17) and then substituting those values in equation (15), we get:

$$SR = \left(\log_2\left(1 + \frac{O_{tx} |H_{i,j}|^2}{\sigma_j^2 + i_p + O_{mc}}\right) - \log_2\left(1 + \frac{O_{tx} |H_{i,in}|^2}{\sigma_{in}^2 + i_p' + O_{mc}}\right)\right)^+ > SR_{th} \quad (18)$$

where, $SR_{th}$ denotes the minimum threshold secrecy rate required to obtain the secured communication. For the case, $SR < SR_{th}$, it can be concluded that the malicious intruder can effectively intrude into the network, $i_p'$ specifies the interference power due to the adjacent users in the vicinity of an eavesdropper. Equation (18) specifies SR to characterize the security of the massive MIMO system. Moreover, this equation can be used to estimate the security performance of the massive MIMO system.

### B. Spectrum Sharing

One of the fundamental constraints that are being targeted with the growing demands of the users is the increased number of spectrum bands. Therefore, the technique of spectrum sharing is introduced. Spectrum sharing technique adapts the primary function to offer efficient service and proficient allocation among multiple users. Considering the scenario of licensed spectrum sharing where sharing players are introduced as homogenous and heterogeneous sharing players. Based on the characteristics of the sharing player, classification of the licensed spectrum sharing scheme is introduced. A homogenous spectrum sharing scheme is the sharing scheme where involved players are of the same nature, while a heterogeneous spectrum sharing scheme involves the spectrum sharing between players of different nature [76]. However, restrictions of the technique affect the security of the network. As LTE was newly projected to operate in an unlicensed band of 5 GHz whose framework coexist with the LSA (Licensed Shared Access) of CBRS for 2.3-2.4 GHz in Europe and 3.5 GHz in U.S. Therefore this technique leads to the increased interference as interference is one key to the factor affecting the QoS and thus decreasing the security efficiency of the network [77].

Congestion is another factor that is required to be examined and counteracted for preventing security attacks, for example, jamming attacks. The taxonomy of threats in spectrum sharing can be categorized into threats possible in sensing driven spectrum sharing and threats possible in database-driven spectrum sharing. Possible menaces to sense driven spectrum sharing include physical layer threats, MAC layer perils, and cross-layer perils. Menaces possible in database-driven spectrum sharing include database inference attacks and a menace to database access protocol involving primary and secondary users [78]. The attacking scenario in the spectrum sharing can be depicted by considering three possibilities. The first possibility involves the secondary transmitter acting as an illegitimate node in spectrum sharing. As shown in Fig. 2, in Example-1 secondary transmitter, behaves as a node of the relay, and therefore supports the primary system transmitter and primary system receiver with the desired data rate.

Considering the secondary system in which secondary transmitter acts as a malevolent node possesses the ability to access information of primary system transmitter and primary system receiver, therefore, have the capability to create a threat on integrity and confidentiality of the information. In Example-2 of Fig. 2., considering multiple secondary receivers as malicious attackers while transmitting secondary data from a Secondary Transmitter (ST) to secondary receiver, there can be the possibility of having multiple valid or invalid secondary receivers. An attack on confidentiality may take place while transmitting information to an invalid secondary receiver. Example-3 shows the scenario of the Primary Transmitter (PT) as an intruder, where the primary system of transmitter and receiver hold the ability to bug the signal of the secondary system transmitter [79].

Several solutions to overcome these attacking threats were executed to fulfill the requirements of security performance. PLS is one of the significant technique that makes use of physical characteristics of the wireless channel to secure wireless communication network against an illegitimate user. A source cooperation mechanism aided with an opportunistic jamming framework in the presence of an intruder aims to provide confidentiality of transmission for the technology of spectrum sharing [80]. Another technique to improve secrecy performance in spectrum sharing includes cooperative relay selection, where three relay selection strategies were followed, namely PRS, MSRS, ORS. ORS is measured as an outperformer in comparison to the other two strategies under the outlay of having full CSI [81]. The primary aspects of the spectrum sharing system involve spectrum sharing, a decision for the spectrum, sharing of spectrum and mobility of spectrum. However, the two major stages of spectrum sharing are mentioned below:

*1) Stage of spectrum sensing:*

Spectrum sensing is amenable for determining the state of the spectrum in the form of idle, busy or available for the secondary access. It is subjected to decreasing the delay, energy consumption, and interference created to the primary system. Irrespective of its advantages, it is associated with certain challenging limitations such as hardware requirements in terms of high-speed processors, high-resolution ADCs (Analog to Digital Converters) in terms of FPGAs, DSPs. Other challenging features include hidden primary user problems, security issues during sensing. One of the major incursions that can have an effect during sensing is the attack on integrity or intervention where integrity is compromised, is known as PUE (Primary User Emulation), where an eavesdropper impersonates the primary system and is capable of deceiving the secondary system. The generalized security modeling of the spectrum sensing can be expressed as:

$$A_0: x(i) = n(i) \qquad (19)$$
$$A_1: x(i) = hy(i) + n(j) \qquad (20)$$

Spectrum sharing is used to determine the state of the primary user by using the above mentioned two conditions, where $i$= 1,2, …, N, N signifies the total number of samples and is the function of sensing time $\mu$ and frequency of sampling $f_N$. $x(i)$ denotes the received signal at secondary user transmitter, $y(i)$ is the transmitted signal by the primary user, $h$ defines the channel characteristics between the primary system transmitter and the secondary system receiver, $n(i)$ depicts the noise.

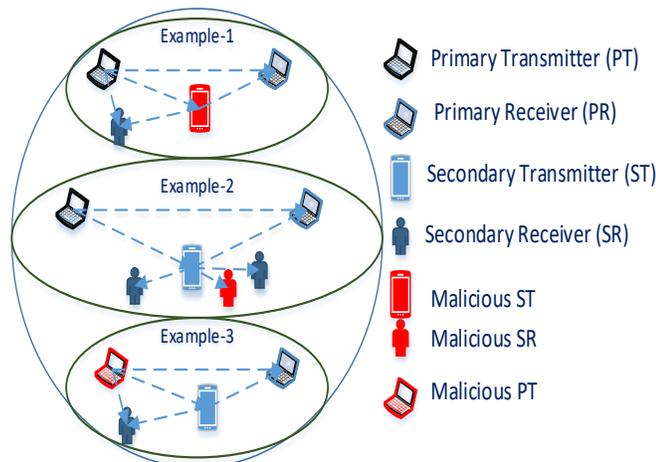

Fig. 2. Possible security attack scenario in spectrum sharing

For sensing case of $A_1$ depicts an active state of the primary user and is termed as perfect detection while as for sensing case of $A_0$, depicting an idle state of the primary user and their corresponding probabilities are known as the detection probability $P(A_1)$ and the false alarm probability $P(A_0)$. The corresponding probabilities of detection and false alarm are expressed as:

$$P(A_1) = Q\left[\left(\frac{th}{\sigma^2} - P - 1\right)\sqrt{\frac{\mu f_N}{2P+1}}\right] \qquad (21)$$

$$P(A_0) = Q\left[\left(\frac{th}{\sigma^2} - 1\right)\sqrt{\mu f_N}\right] \qquad (22)$$

where $th$ denotes the threshold for detection, P denotes the received SNR (Signal to Noise Ratio) at the secondary system transmitter, $\sigma^2$ denotes variance of noise n, Q(.) represents the Q-function.

*2) Stage of spectrum sharing:*

The second stage involved in spectrum sharing is the stage of sharing, as in the scenario of sharing secondary users are capable of having access to the frequency bands which are certified to the primary system only. For the case when the primary communicating link between the primary system transmitter and receiver is opportunistically weak. The secondary transmitter possesses the capability to get connected to the network so that the primary receiver can achieve its desired data rate. The advantage for the secondary transmitter to intervene in the network is that it gets the opportunity for the transmission of its particularized data to the secondary system receiver by utilizing the left out transmit power after connecting with the primary system. Since the data is transmitted in a scenario of a wireless background. Therefore, creates more vulnerabilities for insecure communication. To maintain the security of the spectrum sharing scenario, according to Wyner, "the security of the information is sustained if the authenticated channel is of greater eminence than the eavesdropping channel".

Consider the network consisting of a primary system involving primary system transmitter, primary system receiver secondary system transmitter, and secondary system receiver operating in a full-duplex mode. Considering the attacking scenario in the network of spectrum sharing, where the secondary system is doubtful about the primary system. It involves two phases of transmission included in spectrum sharing. The first transmission phase encompasses the Request to Cooperate (RTC) sent by the primary transmitter to the primary receiver, secondary system. The broadcasted signal by the primary transmitter be $v_{pt}$ [79]. The received signal by the primary system receiver and the secondary system is expressed as:

$$z^{pr} = \sqrt{o}h_{pr}v_{pt} + n \qquad (23)$$
$$z^{st} = \sqrt{o}h_{st}v_{pt} + n \qquad (24)$$
$$z^{sr} = \sqrt{o}h_{sr}v_{pt} + n \qquad (25)$$

where $z^{pr}$, $z^{st}$ and $z^{sr}$ models the received signal by the primary receiver, secondary transmitter, and secondary receiver correspondingly when the signal is broadcasted by the primary transmitter, $o$ is the total power transmitted, $n$ is the AWGN with zero mean presents at the nodes of the primary receiver and secondary system. Based on the strategy of allocation of power, the secondary system transmitter combines the signal of the primary system transmitter $v_{pt}$ with its own signal $v_{st}$ given by:

$$v = \sqrt{o\chi}v_{pt} + \sqrt{o(1-\chi)}v_{st} \quad (26)$$

where $\chi$ denotes the portion of total transmit power for $v_{pt}$ signal, $o\chi$ denotes the portion of the transmitted power of a secondary system for getting the advantage of certified bandwidth from the primary system. $v$ denotes the combined signal that secondary transmitter broadcasts in the secondary and primary system. Thus, the signal broadcasted by the secondary transmitter, received by the primary and secondary system nodes is given by:

$$z^{pr'} = \left(\sqrt{o\chi}v_{pt} + \sqrt{o(1-\chi)}v_{st}\right)h_{pr}' + n \quad (27)$$

$$z^{sr'} = \left(\sqrt{o\chi}v_{pt} + \sqrt{o(1-\chi)}v_{st}\right)h_{sr}' + n \quad (28)$$

$$z^{pt'} = \left(\sqrt{o\chi}v_{pt} + \sqrt{o(1-\chi)}v_{st}\right)h_{pt}' + n \quad (29)$$

The secondary receiver receives the signals both from the primary transmitter and secondary transmitter. The received signal without interference at the secondary system receiver is given as:

$$z^{SR'} = \left(\sqrt{o(1-\chi)}v_{st}\right)h_{sr}' + n \quad (30)$$

The SNR received at the secondary receiver for decoding of the signal $v_{st}$ is given by:

$$P_{sr} = \frac{(a-\chi)o|h_{sr}'|^2}{n} = (1-\chi)Pg_{sr} \quad (31)$$

As the primary transmitter is considered to be a malicious attacker with an objective of attacking the signal $v_{st}$ of the secondary transmitter. The secondary transmitter achieves the SR given as:

$$SR = \{D_{sr} - D_{pt}\}^+ \quad (32)$$

where $D_{sr}$ denotes the achieved data rate at the secondary system receiver from the secondary system transmitter, $D_{pt}$ denotes the achieved data rate at the primary transmitter (malicious attacker) from the secondary transmitter and are given by:

$$D_{sr} = \frac{1}{2}\log_2(1 + P_{sr}) \quad (33)$$

$$D_{pt} = \frac{1}{2}\log_2(1 + P_{pt}) \quad (34)$$

The (1/2) factor represents the two phases of transmission. For two phases of transmission, the signals are required to be transmitted in two-time slots. Form the above equations, it is observed that security has a direct impact on the data rate achieved at the secondary receiver and on the data rate achieved by the malicious attacker. Form equation (32), the secrecy rate tends to reduce with an upsurge in the rate of data achieved by the malicious attacker. In other words, the security of the data is said to be satisfied if the data achieved by the secondary transmitter is higher than the data achieved by the intruder. It indicates the eminence of the wiretap channel, from secondary system transmitter to primary system transmitter (intruder) not as much that of the quality of the main channel, i.e., from secondary system transmitter to secondary system receiver.

*C. Beamforming*

It is a propitious technique for next generation WCN. It involves the maximization of SNR in a definite direction and minimizing it in former directions resulting in the formation of a beam in a specific direction, which leads to an increase of energy efficiency as the energy is focused instead of wide-ranging [82]. Therefore, provides better QoS [83]. It is significantly convenient in the field of communication and radar. Further, this concept can also be utilized to ameliorate the quality of localization of UE [84]. However, with its numerous advantages, there are certain limitations associated with it. Digital Signal Processing (DSP) chip design is one of the main challenges in adapting the beamforming approach. Power requirements in the beamforming technique are relatively more as compared to other promising techniques of the future generation. Another aspect that affects the performance of the network is the parameter of a battery lifetime, which is estimated to be comparatively lesser. These challenging parameters ultimately affect the security of the WCN.

Although directive beamforming is considered as one of the conceivable tactics to enhance the security performance of the wireless communication network, due to sharing of information in distributed and collaborative beamforming issue of security could be a matter of concern. Both the cases of beamforming, including distributed and collaborative beamforming. It involves the stage of sharing of information, from the source node transmission, which will be perceived by all immediate nodes. The nodes that are not involved in cooperating can observe the same information. It, creates a threat to the integrity of information [85]. The standard of IEEE 802.11ad identifies a selection based protocol for the technique of beamforming. It involves three phases. The first phase is the phase of SLS involving quasi Omni-pattern to choose the best transmit and receive antenna sector. The second phase involves the phase of beam refinement to select better beamwidth beam pattern pair. This phase is followed by the optional beam tracking phase, where channel changes are adjusted during data transmission [86]. The methods of beamforming encompass ZF, MRC, and MMSE [87]. Mathematically beamforming can be represented as:

Consider a multiple antenna system with $j$ receive antennas with $o_1, o_2, \ldots, o_J$ as $j$ receive symbols at different receive antennas and single transmitting antenna with $m$ as the transmitted symbol $p_1, p_2, \ldots, p_J$ as the AWGN with zero mean and $h_1, h_2, \ldots, h_J$ represent channel coefficients. The general system model can be expressed as:

$$\bar{o} = \bar{h}m + \bar{p} \ldots \quad (35)$$

On applying beamforming, represented by the weighted combination of received symbols as:

$$s = [w_1 \ w_2 \ w_3 \ \ldots \ w_j]\begin{bmatrix} o_1 \\ o_2 \\ \vdots \\ o_j \end{bmatrix} \quad (36)$$

It can also be denoted as

$$r = \bar{w}'\bar{o} \ldots \quad (37)$$

where $\bar{w}$ is the vector of weights and is also termed as a vector of beamforming. Therefore, by choosing optimal weights, a beam is steered in a specific direction and thus suppressing the beam in other directions. The process of finding the optimal value of $\bar{w}$ is known as the beamforming problem. Beam steering is executed by a process known as electronic steering offers low complexity, increased energy efficiency, and enhanced precision. For adapting $\bar{w}$ and obtaining the optimal weight vector from the equation (37) as :

$$\bar{w}'\bar{o} = \bar{w}'\bar{h}m + \bar{w}'\bar{p}\ldots \quad (38)$$

where $\bar{w}'\bar{h}$ depicts signal gain and $\bar{w}'\bar{h}\bar{p}$ represents the noise present at the output of beamforming

However, SNR can be maximized by keeping the signal gain constant and minimizing the noise power such that:

*1) Case 1:*

$$\bar{w}'\bar{h} = 1 \ldots \quad (39)$$

Or
$$\bar{h}'\bar{w} = 1 \ldots \quad (40)$$

Equation (39) represents the equation of constraint for the optimization problem, where all vector $\bar{w}$ lie on hyperplane and indicates a unit gain in signal direction. Formulating SNR maximization forms the objective function for the optimization problem. From equation (38), noise power can be specified as:

$$\bar{w}'\bar{p} = \sum_i w_i p_i \quad (41)$$

An assumption is made where all noise samples are statistically identical with zero mean and identical variance $\theta^2$ signifying IID (Independent and Identically Distributed) such that

$$E[(\sum_i w_i p_i)^2] = \sum_i \theta^2 w_i^2 \quad (42)$$

The above can also be represented as:

$$E[(\sum_i w_i p_i)^2] = \theta^2 ||w||^2 \quad (43)$$

The optimization problem for beamforming involving minimizing the noise power subject to the constraint $\bar{w}'\bar{h} = 1$, indicating the constant signal gain

*2) Case 2: if $\bar{w}'\bar{h}$ is an affine constraint and the objective is convex gain as:*

$$\theta^2 ||w||^2 = ||w||^2 \ldots \quad (44)$$

Equation (44) indicates the convex objective. Both convex objective and convex constraints constitute convex optimization problem and to solve a convex optimization problem LaGrange's multiplier can be operated, represented by:

$$L = \bar{w}'w + \zeta(1 - \bar{w}'\bar{h}) \ldots \quad (45)$$

Equation (45) indicates the use of LaGrange's multiplier denoted by $\zeta$ and the equation for optimal beamformer can be given as:

$$\bar{w} = \left(\frac{\zeta}{2}\right)\bar{h}\ldots \quad (46)$$

Equation (46) indicates the optimal beamformer with maximizing the SNR ratio. It is also be depicted that $\bar{w}$ is proportional to $\bar{h}$ which is analogous to the spatially matched filter. However, to determine $\zeta$, the constraint $\bar{w}'\bar{h} = 1$ is used. Therefore, $\zeta$ is calculated as:

$$\zeta = \frac{2}{||\bar{h}||^2} \quad (47)$$

Using the above equation optimal beamformer $\bar{w}^*$ is calculated as:

$$\bar{w}^* = \frac{\bar{h}}{||\bar{h}||^2}\ldots \quad (48)$$

Equation (48) is also termed as the maximal ratio combiner as it maximizes SNR employing beamforming and the SNR can be given by:

$$SNR = \frac{E[m^2]}{\theta^2 ||\bar{w}||^2} \quad (49)$$

For any combiner SNR can also be written as:

$$SNR = \frac{a}{\theta^2}||\bar{h}||^2 \ldots \quad (50)$$

The above equation (50) can also be represented in terms of transmit power is given by:

$$SNR = \varrho ||\bar{h}||^2 \quad (51)$$

where $\varrho$ symbolizes the transmit power and is interpreted as the ratio of signal power and noise power. SNR has a definite impact on the security of the WCN, as the quality of main channel increases with an increase in SNR and decreases the interferences present in the channel, which in turn increases the channel capacity as shown in equation (16) and (17), thus affects secrecy rate. From equation (15), the secrecy rate is precisely explained as the difference between the capacity of the main channel and capacity of the channel of the malicious observer, determining the security of the WCN. From equation (48) and (50) by obtaining optimal beamformer SNR can be maximized. It gives rise in capacity of the main channel and the reduction of the interference resulting in improved performance of the security.

*D. Device to Device (D2D) communication*

The 5G offers device centric architecture of network in comparison to previous generation network architectures, and a device is expected to perform dynamically. Therefore, defining D2D communication as the shortest communication between device nodes without the need to pass through the core network or through an access point [88]. Thus, without the use of central authority, including base stations and access points, the functions of logging, auditing are controlled by resource-constrained end-user devices. Therefore, D2D communication executes a crucial role in next generation WCNs. It provides the ability to offload heavy data traffic by improving the utilization of network resources, provides short-range communication, thus supports proximity based services. Also, D2D communication depends on the device discovery to perceive peers of communication, which is performed by transmitting information over wireless channels [89]. In spite of its numerous advantages, several security threats are emerging due to its wireless broadcast communication nature, which allows malicious attackers to track and locate D2D users, creating a threat to location privacy. The users of D2D are spontaneous and self-managed, enforcing and managing security is still a challenge as compared to the centralized cellular network. Moreover, next generation WCN architecture and latest application scenarios expose D2D communication to eavesdropping, jamming, privacy violation, data modification. Mainly D2D communication adopts the connections between devices, which creates a menace to security due to the direct wireless connection, end-user mobility, and privacy issues. With an increasing number of devices adapting D2D communication, there is an increase in adversaries to target the security of the D2D WCN [90].

TABLE I
TECHNIQUES TO ELEVATE THE SECURITY OF THE 5G NETWORK

| S. no. | Technique | Objective | Description | Outcomes/ results |
|---|---|---|---|---|
| 1. | Prisoner's dilemma game theory [12] | Analysis of Bandwidth spoofing attack on various stages of WCN. | Approach to evaluate the victory of the intruder through bandwidth spoofing. | Intruder capable of spoofing bandwidth with a significant winning percentage from a legitimate user. |
| 2. | Intrusion detection technique using Adaptive Neuro-Fuzzy Inference System (ANFIS) [14] | Detection of an intruder at relay using ANFIS | Intruder detection system ANFIS providing minimum average testing error and the malicious attackers are identified on the basis of the predetermined threshold value | ANFIS provides minimum average testing error which is being chosen as a threshold, and therefore the detection of an intruder is done based on the attained value of the threshold |
| 3. | Physical layer security (PLS) technique [21] | Improving the security of the heterogeneous network | Spatial modeling, mobile association are the PLS techniques incorporated to enhance the security | Secrecy rate tends to increase profoundly by incorporating beamforming with artificial noise (AN) |
| 4. | Linear precoding and Artificial Noise (AN) technique [37] | Secrecy enhancement using linear precoding of information and by utilizing AN | Downlink for massive MIMO multi-cell incorporating linear precoding and AN for enhancement of physical layer security | Ergodic secrecy rate is analyzed in the presence of contamination of the pilot signal by using AN and linear precoders. |
| 5. | Thermal pattern analysis technique [100] | To identify the most probable area of illicit access for high-speed handover setup | Depending upon the power consumed by user equipment at any instant and tracing its energy pattern, detection of the low security region is observed | Regions with low security can be traced out by comparing the real-time energy requirement estimates of devices |
| 6. | Secure harvesting technique [101] | Safeguard of extremely penetrating zones, where the possibility of invalid access is maximum using RBA | Highly probable regions of attack are being protected by implementing RBA using encryption. | PDF (Probability Density Function) detection framework is analyzed to protect the poorly secured zones |
| 7. | Boundary technique [112] | To improve the PLS by analyzing the receiver performance | The probability of secure connectivity is calculated using Wyner coding scheme. The procedure involves the hiding of receiver present at the corner. | The secrecy rate for a high data rate is observed to be increased by hiding the receiver at the boundary or corners. |
| 8. | Reinforcement learning technique [113] | Secure offloading to the edge nodes along with a collaborative caching scheme for data privacy | Learning agent is optimized by using trial and error to obtain the optimal resultant strategy | A security solution to provide protection against numerous smart attacks in mobile edge caching systems |
| 9. | Multi-antenna transmission technique [116] | To ameliorate the physical layer security in Millimeter-Wave Vehicular Communication | Use of multiple antennas utilizing a single radio frequency chain for the transmission of symbols of information to the destined receiver and noise to non-receiver | Higher secrecy rates are observed in comparison to physical layer security techniques using digital or complex antenna architectures |
| 10. | Network function virtualization technique for Botnet Detection and mitigation [120] | To detect and reduce botnets | It involves high-level detection phase by flow monitoring function and Deep Packet Inspection (DPI) for the configuration of the detected bot | Personalized and virtualized honeynet for the proactive detection and mitigation of botnets |
| 11. | Beamforming and jamming technique [124] | To offer PLS by using AN (Artificial Noise) aided the beamforming system | An AN is acting as a jamming signal to obstruct malicious attackers. | the expected SR loss is linearly rising with respect to the power dividing factor and to the numeral of transmit antennas and shows constant behavior with respect to the numeral of malicious intruders. |
| 12. | Physical Layer authentication technique [130]. | To use multiple landmarks with multiple antennas for the improvement of spoofing detection accuracy | Multiple MIMO landmarks are used to determine the signal RSSI (received signal strength indicators) with a logistic regression approach | Improved accuracy of spoofing attack detection with reduced overheads |
| 13. | Robust allocation of resources technique [145] | To enhance the PLS in the presence of an active intruder | Robust covariance matrices of transmitting end are obtained for the scenario of Half Duplex legitimate receiver friendly jammer (HDJ), Half Duplex legitimate receiver (HD), half-duplex legitimate receiver (Half Duplex legitimate decode-and-forward relay-HDR), full-duplex (FD) scenarios | Better performance is achieved if the jammer is placed close to the malicious intruder. |
| 14. | Diversity technique , [146] | To upsurge the protection mechanism of the physical layer | Security performance is evaluated for cooperative relay present in an environment pertaining to Rayleigh fading in terms of secrecy capacity and the probability of interception | Security is observed by growing the number of cooperative relays using the scheme of a better selection of relay |

Table I illustrates the brief summary of various techniques that have been experienced in improving the security of the 5G communication network. These techniques form the basis and lay the foundation for improving security.

However, the proper mechanism of security is still required to be modeled to enhance the capability of secure integration with future generation architecture and its applications. Through various research has been studied for full filling the security parameters; however, an effectual mechanism of the security solution is still a challenge in the field of research. One of the threats to the security in the D2D communication network is the proximity based malicious attack where a mobile UE while acting as a D2D server can get proximity based mobile viruses and are not therefore, able to offer service to other devices. In addition to it, compromised security of UEs further become a new source of threat to security while communicating with other devices and thus consequently can spread the attack to the entire network. For mitigating such threats, one approach involves the utilization of non-cooperative game theory and epidemic theory to understand the interdependent security risks and enables security-aware incentive mechanisms [91].

Another approach of enabling a security-enhanced mechanism for cellular users in D2D involves the combined optimization of channel assignment and power allocation of communicating D2D links where downlink resource sharing is used via multi-channel and single channel communication. On the basis of the optimized D2D data rate in view of the power budget, further improvements in security are made accordingly [92]. Additional efforts that were made to secure D2D communication include pairwise establishment of key to introduce short authentication based string agreement protocol [93]. The technique of link adaptation to create a balance between secrecy spectral efficiency and secrecy energy efficiency [94]. The PLS technique were adapted to analyze the rate of generation of secure key on the basis of cooperative relays (trusted and non-trusted) [95]. The power transfer model is evaluated for the transmission of secured information by making use of three defined policies: best power beacon, cooperative power beacon, and nearest power beacon [96], combined scheme of public-key cryptography and mutual authentication [97]. The generalized security model of D2D communication:

Consider the scenario of a D2D communication network, where a device $D_1$ is far from receiving device denoted by $D_2$ and it is not possible for the device $D_1$ to directly communicate with the receiving device denoted by $D_2$. Consider another device $D_r$ present in the network in such a manner such that it is present in connectable premises with both devices and therefore, can act as a relay device. Consider M as an eavesdropper under the assumption that the CSI of the uplink is known and the eavesdropper attempts to obtain the information that the devices $D_1$ and $D_r$ communicates. Communication follows two phases. In the first phase, BS makes the decision of allocating the RB (Resource Block) to the D2D communication, and in the second phase relay device makes the transmission of the signal received from the first and its own information to the receiver device. The channel is modeled under the Rayleigh fading such that the channel gains on the resource block are given by $h_1^c, h_2^c, h_3^c, \dots, h_n^c$ for the communication links, $h_1^i, h_2^i, h_3^i, \dots, h_n^i$ as interference links and $h_1^M, h_2^M, h_3^M, \dots, h_n^M$ as eavesdropper links, such that for the first phase at the BS the received signal, $D_r$ and eavesdropper M respectively is given by:

$$R_j = \sqrt{o_c} v_c h_C^j + \sqrt{o_1} v_1 h_1^j + n_j \quad (52)$$

where j denotes BS, the relay device $D_r$ and eavesdropper M, $o_c$ and $o_1$ depicts the power of cellular user in transmit mode and device $D_1$, $n_j$ depicts the AWGN with zero mean and variance $\sigma^2$, $h_C^j$ represents the channel coefficient form cellular user to corresponding $J$ = BS, a relay device, eavesdropper, $h_1^j$ represents the channel coefficient form transmitting device $D_1$ to corresponding BS or relay device or eavesdropper, $v_c$ and $v_1$ represents the transmitted information from the cellular user and from device $D_1$ respectively. From above the instantaneous SINR at the BS, $D_r$ and at the eavesdropper m is given by:

$$P_j = \frac{o_c \left|h_c^j\right|^2}{o_1\left|h_1^j\right|^2 + \sigma^2} \quad (53)$$

For the second phase the received signal at the BS, eavesdropper M and at the receiver device $D_2$ is given by:

$$R'_{BS} = \sqrt{o_c} v_c h_C^{BS} + h_2^{BS}\left(\sqrt{lo_1} R_{D_r} + \sqrt{(1-l)o_{D_r}} v_{D_r}\right) + n_{BS} \quad (54)$$

$$R'_2 = \sqrt{o_c} v_c h_C^2 + h_2^2\left(\sqrt{lo_{D_r}} R_{D_r} + \sqrt{(1-l)o_{D_r}} v_{D_r}\right) + n_2 \quad (55)$$

$$R'_M = \sqrt{o_c} v_c h_C^M + h_2^M\left(\sqrt{lo_{D_r}} R_{D_r} + \sqrt{(1-l)o_{D_r}} v_{D_r}\right) + n_M \quad (56)$$

where $l$ depicts the cooperation level, $v_c$ and $v_{D_r}$ represents the transmitted information from cellular users and from the relay device $D_r$, $n_{BS}, n_2$ and $n_M$ depicts the circularly symmetric AWGN with zero mean and variance $\sigma^2$ at the BS, device $D_2$ and at the eavesdropper. Therefore, the instantaneous SINR obtained at the BS, device $D_2$ and an eavesdropper M is given by:

$$N_{BS} = \frac{o_c\left|h_C^{BS}\right|^2}{o_{D_r}\left|h_2^{BS}\right|^2 + \left(l\left(o_1\left|h_1^{BS}\right|^2 + o_c\left|h_C^{BS}\right|^2 + \sigma^2\right) + (1-l)\right) + \sigma^2} \quad (57)$$

$$N_2 = \frac{o_{D_r}\left|h_C^2\right|^2\left(lo_1\left|h_1^{D_r}\right|^2 + (1-l)\right)}{o_c\left|h_C^2\right|^2 + lo_{D_r}\left|h_1^2\right|^2\left(\sigma^2 + o_c\left|h_C^2\right|^2\right) + \sigma^2} \quad (58)$$

$$N_M = \frac{o_{D_r}\left|h_C^M\right|^2\left(lo_1\left|h_1^{D_r}\right|^2 + (1-l)\right)}{o_c\left|h_C^M\right|^2 + lo_{D_r}\left|h_1^M\right|^2\left(\sigma^2 + o_c\left|h_C^M\right|^2\right) + \sigma^2} \quad (59)$$

Therefore the data rate for the scenario where the communication takes place through the relay device such that, if c number of channels are reused such that $c < n$, for the D2D link, the data rate on channel c is given by:

$$D_c \triangleq \log_2\left(1 + \frac{o_c^x g_x^c}{1 + o_c g_c^{x'}}\right) \quad (60)$$

where $D_n$ denotes data rate on channel n of D2D link $x$ $n$, $o_c^x$ is the power transmitted with D2D link $x$ on channel c, $g_x^c$ and $g_c^{x'}$ are the normalized gains and are expressed as:

$$g_c^{x'} \triangleq \frac{|h_c^{x'}|^2}{\sigma_{x,c}^2} \quad (61)$$

$$g_M^{c'} = \frac{|h_M^c|^2}{\sigma_{M,x}^2} \quad (62)$$

$$g_c^{M'} = \frac{|h_c^M|^2}{\sigma_{M,x}^2} \quad (63)$$

$$g_x^c \triangleq \frac{|h_x^c|^2}{\sigma_{c,x}^2} \quad (64)$$

Where $h_c^x$ is the channel coefficient on channel c from D2D link x transmitter to its respective receiver, $h_c^{x'}$ is the cross channel coefficient from the BS to D2D link $x$ receiver. Therefore the secrecy rate can be achieved as:

$$D_{c,M} \triangleq \left[ \log_2\left(1 + \frac{o_c g_x^c}{1 + o_c g_c^{x'}}\right) - \log_2\left(1 + \frac{o_c g_M^{c'}}{1 + o_c g_M^{x'}}\right) \right]^+ \quad (65)$$

From the above equation achieved, it is clarified that the achieved rate of secrecy is dependent on the normalized gains and their respective powers [151]. Where normalized gains are in turn, the function of channel gains depicting an improved gain of the main channel without intruder improves the performance of the security.

*E. UDN (Ultra Dense Network)*

UDN is investigated as the technology capable of justifying the growing demands of data rate, low latency, seamless connectivity, ubiquitous coverage. For the fulfillment of extensive applications of existing generation and next-generation WCN, UDN supports a large increase in a number of small cells (macro, pico, femto) with increased density of small cell access (SCA) points. There are certain challenges associated with it, such as small cell discovery, user association, energy efficiency, backhauling, propagation modeling, interference management [98]. However, it exposes more vulnerable breaches for an attacker to attack and decrease the security of the network. Various possibilities of invalid access in UDN and other security challenges associated with it are mentioned in [99]. Handover management attacks are expected to be considered with more frequent possibility, as the figure of handovers is instinctively increased due to the improved number of small cell densities. Various approaches have been investigated for the management of handover attacks such as RBA, TPA, but an efficient security mechanism is quite devoid of being contented [100],[101].

Several approaches were followed to improve the performance of the security in UDN. PLS approach is used in UDN while considering the density of users and the density of eavesdropper. Their impact on the average secrecy rate following the method of stochastic geometry to obtain the secrecy rate [102]. A mechanism of blockchains is employed to provide the authentication of the network. [103] A comprehensive overview of efficient solutions is given in [104], [106] to improve physical layer security. An approach of BS densification is considered as one of the approaches to increase the security of UDN, as with the growth in the number of BS. Consequently, gain tends to increase due to which path loss gets decreased. However, this phenomenon is also associated with frequent handovers and increased inter-cell interference from neighboring active BSs. The large antenna array is another tactic to improve physical layer security. BSs employed with a large number of arrays of an antenna can significantly provide large gains to the valid users and therefore improves the performance of secrecy. In case of heterogeneous UDN, mm-Wave and sub 6 GHz BSs are capable of providing varying levels of array gains depending on the precoding and beamforming designs, therefore degrades channels of eavesdropping

PLS has referred to the core idea to improve security performance for the densified networks. Resource management is one of the approaches to enhance security in a UDN. Another approach towards an increase in security is the blockage of an eavesdropper [105]. Physical layer security can also be upgraded by considering the UDN, where users are examined under the close proximity to the cells. The effects of user density, cell density, and eavesdropper are observed on the average SR. Considering the stochastic geometry under the Rician fading channel, an expression for average SR is determined for the wiretap channel and the main channel without leakage [107]. Additional access to improve the security of the network is user-centric UDN by utilizing the methodology of the network providing service to user and de-cellular. The secure service of the network is provided by the dynamic AP (Access Point) grouping enabling mobility management, interference management, resource management, and most importantly security issues [108]. Another user-centric approach involving clustering is followed for UDN occupying the secrecy and energy efficiency perspective. A dedicated jamming strategy, along with an embedded jamming strategy, is followed to satisfy the security requirements of UDN. This design is formulated for both unknown and known channel state information. A set of empirical greedy protected user-centric clustering algorithms is operated for different scenarios of the UDN [109].

In the case of the machine to machine communication in UDN, the vulnerabilities considerably exist more due to the cellular network of the UDN. Also, the high density of BSs creates the issue of frequent handovers and hence, more vulnerable sites for an eavesdropper [110]. However, various categories of attacks and their source causes are stated in [111]. A technique of secrecy improvement is given in [112], where the average capacity of secrecy is evaluated on the basis of Poisson interferers. Another technique is contrary to a jamming attack includes the adaptation of reinforcement learning to provide secured offloading [113]. For high-speed users in UDN under the scenario of dense picocells, where users are considered to be randomly distributed, the issue of security for the framework of vehicular users in UDN is explained by the technique of BM and BB. This technique improves the secrecy of communication by improving the reliability of the information exchange between source and destination. The parameter of FSA (Fairness Security Assessment) is analyzed to determine the potential capability of BM and BB and hence improved the value of secrecy rate [114]. A scheme of mobile association based on the threshold of the received signal power is modeled in [115] to improve the security of the network. The use of millimeter waves in UDN is considered to be more secure and more suitable for the transfer of power wirelessly, as the path loss incorporated in short distance is lesser as compared to the path loss for long-distance communication[116]. The effective secrecy throughput in the presence of multiple malicious observers is analyzed, and protected communication for SWIPT is observed [117], [118]. Also, in [119], Matern Hard Core Point Processes is used to characterize the haphazard position of the BS and users and thus determine the secrecy outage, respectively. The generalized security model of UDN communication is expressed as:

Consider the scenario of UDN consisting of small cell base station transmitting a power '$P_{bs}$' where the position of the base station is spatially distributed according to homogenous Poisons Point Process (PPP) $\Gamma_{bs}$ with a density of $\Psi_{bs}$ and another homogenous PPP of mobile stations $\Gamma_{ms}$ with a density of $\Psi_{ms}$ held by valid users independent from BSs PP. Consider a malicious attacker distributed according to homogeneous PPP $\Gamma_{ev}$ having a density of $\Psi_{ma}$ and is independent of PPPs of BS and valid users, present along with the valid users, and attempts to intercept the communicating data of valid users.

Let the density of valid users is less than the density of malicious attackers such that $\Psi_{ev} < \Psi_{ms}$. Also, channel characteristics of small scale fading are considered adapting the fundamental property of UDN, where LOS (Line of Sight) is the most possible propagation. The main channel is characterized as the channel between the valid users and base stations denoted by $h_m$ while as the eavesdroppers channel or the leakage channel is characterized as the channel between the malicious attacker and transmitting BS denoted by $h_{ev}$. To evaluate the secure communication in UDN, the parameter of secrecy rate is examined denoted by $S_r$ is expressed as:

$$S_r = [S_m - S_{ev}]^+ \quad (66)$$

where, $S_m$ is the secrecy rate of the main channel and $S_{ev}$ is the secrecy rate of the channel of an eavesdropper, where $[a]^+ = \max[a,0]$ and the average secrecy rate can be estimated as:

$$\bar{S}_r \approx \bar{S}_m - \bar{S}_{ev} \quad (67)$$

where $\bar{S}_r$ is the average secrecy rate, $\bar{S}_m$ is the average main channel secrecy rate and $\bar{S}_{ev}$ is the average secrecy rate of the channel of the eavesdropper

For the Eavesdropper channel or the link with leakage, where the nearest malicious attacker is considered as the most damaging malicious attacker [107] and the average channel secrecy rate with leakage in the presence of the nearest eavesdropper where other eavesdroppers are independent of each other is given by:

$$\bar{S}_{ev} = 2\pi \, \Psi_{ev} \int_0^\infty b e^{-\pi \Psi_{ev} b^2} E\left[\ln\left(1 + \frac{Ph_{ev}b^{-\gamma}}{\sigma^2 + F_{ev}(b)}\right)\right] db \quad (68)$$

where E[.] denotes the operator of expectation, $h_{ev}$ denotes the channel characteristics of the leakage channel of eavesdropper and b is the distance between BS and nearest eavesdropper and $F_{ev}(b)$ is the interference present in the leakage link between BS and eavesdropper and is given by:

$$F_{ev}(b) = \sum_{c \in \Gamma_d/c_o(b)} Ph_{ev}b^{-\gamma} \quad (69)$$

where $\Gamma_d$ are the thinned small cells point process $\Gamma_d$ is the subset of PPP of BS.

$$\bar{S}_s = 2\pi \, \Psi_s \int_0^\infty a e^{-\pi \Psi_s a^2} E\left[\ln\left(1 + \frac{Ph_s a^{-\gamma}}{\sigma^2 + F_s(a)}\right)\right] da \quad (70)$$

where $h_s$ is the gain of the main channel, a is the distance between the nearest BS and the specific user, $F_s(a)$ is the aggregate interference present in the main link due to other active base stations and is given by:

$$F_s(a) = \sum_{c \in \frac{\Gamma_d}{c_o(a)}} Ph_s a_c^{-\gamma} \quad (71)$$

From the above equations, it can be concluded that nearest intruders decrease the secrecy rate more effectively as compared to intruders that are far from the respective Base Station, thereby creates a primary impact on the security of the communicating network.

*F. IoT (Internet of Things)*

IoT describes the future generation epoch of wireless communication providing immense convenience and value. It is capable of connecting the number of entities, including sensors, mobile phones, automation, smart vehicles, smart home appliances, on-body sensors in the scientific and industrial community. In general, both the survival of humans, as well as industries, are drastically affected by the massive deployment of application critical IoT devices. It has become the collective term for upcoming networks [121]. From preceding years, the foundation of undefended project masses created vulnerable sites causing numerous "proof of concept" attacks to exploit those exposures [122]. Therefore, with the rapid deployment of such a promising era, it carries certain challenges with it in terms of security management, energy efficiency, autonomy, scalability, and other performance tradeoffs. The cause of these challenges is easy to follow, as the information exchange is now much more life-critical and sensitive.

Personal health monitoring, biometric data, information of location are present all over the internet, and the possibility of breaches could create the susceptibility of organized e-crimes resulting in enormous financial the damages, the depiction of reputation of the company to stakeholders and customers, and more importantly, creates a threat to the lives of human [123]. Therefore, among these challenges, security occupies the critical priority for the secure communication irrespective of its distributed and broadcast nature of communication. As the internet of things is centered by the number of seamless connections and generally IoT architectures encompass four layers viz network layer, perceptual layer, application layer, and the support layer. The perceptual layer is the primary layer responsible for collecting information of all types from the physical world by making use of devices, including radio frequency identification tags or accelerometers.

On the basis of simplicity and resource-oriented nature of the IoT entities present at the perceptual layer, the security of IoT communicating signals at this layer is a challenging task. Various security solutions were proposed for the authentication of IoT signals such as traditional encryption techniques, PLS techniques to secure communication. A lightweight biometric system specifically considered for limited resources of IoT entities is proposed in [121] for the authentication of the user. On the basis of the MCC code, which is basically a state-of-the-art matching fingerprint algorithm, it raises the security of the network by maintaining the accuracy of recognition. The biometric system makes use of block logic operation, which results in a decrease of biometric feature size, therefore reduces the number of computations and memory.

In [123], a PUF based cryptosystem design was proposed for the lightweight security of IoTs. This cryptosystem is based on the principle that two chips that are identical from the same line of production cannot share the alike and exact physical characteristics. These characteristics may include frequency of oscillating components, speed of racing signals, the randomness of memory components in their initial states, speckle outline and arrangement of optically delicate substantial. On the basis of such characteristics, devices for cryptographic functions can be mechanized in the form of PUFs. A mechanism of PLS enhancement in IoTs is proposed in [126], where the performance of security is evaluated for a multiuser system that makes use of antenna selection scheme at the transmitter end TAS and incorporates diversity opportunistic scheduling TSD on the basis of threshold over legitimate nodes. The valid communication between the legitimate source and authentic receiver experiences the existence of colluding and non-colluding passive malicious attacker(s). Closed system expression for PDFs and CDFs (Cumulative Density Functions) of SINR (signal-to-interference-plus-noise ratio) and other evaluating parameters of security such as Secrecy Outage Probability (SOP) of the system, asymptotic outage secrecy probability for the scenario of colluding and non-colluding eavesdroppers were observed.

Another approach to secure communication in the internet of things involves a watermarking algorithm [127]. In such an approach, dynamic authentication is performed for the detection of cyber-attacks. This approach is based on a deep learning mechanism to provide short term memory structure in order to abstract a determined set of probabilistic components of the produced signal and therefore enabling the entities of IoT to dynamically watermark the features of the set into the signal. The authentication process is effectively performed by enabling the IoT gateway to collect signals from the devices of IoT to verify the reliability of the signal. In [128], PETs were studied where state-of-the-art principles for the laws of privacy, considered for the architecture, were analyzed. The valid principles for the implementation of PETs at various layers of IoTs were demonstrated. Privacy legislation maps validated with respect to the principles of privacy are employed for the design of PETs in the architecture of IoT stack.

A deep neural network that permits authentication of wireless nodes in real-time on the basis of variation in properties of RF of the transmitter detected by in-situ ML present at the destination end [129]. This scheme makes use of 65-nm node standard, including factors such as local oscillator offset, I-Q imbalance where detection using neural network consists of the hidden layer having 50 neurons and is capable of providing accuracy of 99.9% with an ability to distinguish 4800 transmitters. It is responsible for providing authentication under varying channel conditions. A mechanism of spoofing attack detection by operating the system of PL authentication by exploiting the CSI with more accuracy using multiple landmark schemes [130]. The risks that lie in the privacy and security of IoT are delved in [131].

In [132], the IoT attack model was identified based on the learning approach to improve security in terms of secure offloading, access control, authentication, malware detection. For malware detection, an approach of Q-learning is followed to obtain optimal offloading rate with unknown bandwidth model and trace generation of the neighboring devices of IoT. For secure offloading, the technique of reinforcement learning is approached in the dynamic radio environment. For access control, the machine learning approach, including techniques such as K-NNs (K-Nearest Neighbor), SVMs, and NNs, are capable of detection of an intruder in IoTs. Moreover, another field where the security of IoTs occupies fundamental importance in the field of healthcare. In the case of the smart healthcare system, various security issues exist in the form of password guessing attacks, which can lead to DoS attacks, active and passive attacks, respectively, creating a menace to the privacy of health data. Keeping in view the concept of a learning approach, IoT security challenges are required to be addressed proactively where the devices can adapt dynamically security threats and possess reconfigurability [133].

Various approaches were followed to increase the strength of the password and therefore reduce the security vulnerabilities. PSMs approach is proposed in [134] to progress the security of healthcare system in IoT on the basis of personal identification information, and thereafter label processing to detect the personal information present in the password and other hidden variations. The GSC and RSC are the two antenna reception techniques that are used to improve the security on the basis of the zero-forcing beamforming algorithm [135]. One more approach to enhance the security of the IoTs is to secure the downlink of the transmission by using cooperative jammer against non-colluding and passive eavesdroppers, where analysis of security parameters such as SR and SOP are analyzed [138]. The emerging challenges in the security of IoTs whose countermeasures are yet to be solved are an exponential rise in the numeral of poor links and unpredicted use of information [139]. The significant areas that are present in the current IoT landscape include security in connected vehicles (VANETs-Vehicular Ad-hoc Networks), security in MEC (Mobile Edge Computing), blockchains, artificial intelligence as a solution to enhance the security in the IoTs [140], [142].

A polar low complexity coding method is utilized to provide confidentiality to the information under improved secrecy [141]. Blockchain, as a possible solution to the security of IoT, acts as a decentralized database on the basis of cryptographic techniques to improve the security of the IoTs and, therefore, provides reliability, desirable scalability, and authentication [143]. However, conventional mechanisms to improve the security of IoT requires an optimization framework for the involved nodes in the network to provide valid access to nodes [149]. Also, a countermeasure to the Sybil attack involves K-means clustering, followed by the detection scheme via age replacement policy is identified [150]. Yet, It involves the examination of only one intruder node. Considering all these above mentioned security enhancement schemes, a novel and sophisticated procedure of filtering, firewalls, potent IDS are required to enable secure health care, business practices, and other essential applications. The mathematical security model for IoTs can be evaluated as:

Consider a system with n number of users where n≥1 such that a scenario of the multiuser downlink is obtained. Assume co-channel interference is present, and each user (valid and eavesdropper) suffers identical co-channel interference. Consider A number of antennas at the transmitter end and B number of users, which are involved in communication from a set of n number of users. Consider the presence of an eavesdropper either in an active or passive attacking mode, possessing the capability to intercept the communicating information. In addition to it, assume the communication link is I.I.D. The expected signal at the $m^{th}$ user at the $p^{th}$ antenna BS is derived as:

$$R_{m,p} = \sqrt{c_m} h_{m,p} d + \sum_{i_p=1}^{I_p} \sqrt{c_{i_p}} h_{i_p,p} d_{i_p} + n_m \quad (72)$$

where $c_m$ is the power transmitted to the valid user, $h_{m,p}$ characterizes the channel coefficient between the $m^{th}$ antenna and $p^{th}$ valid-user where m ranges from 1 to A and p ranges from 1 to B, d is the symbol transmitted to the valid user having zero mean and unit variance, $c_{i_p}$ is the interference power at the $p^{th}$ valid user, $h_{i_p,p}$ is the channel coefficient between $i_p$ interferer and $p^{th}$ valid node, where $i_p$ ranges from 1 to $I_p$, $d_{i_p}$ are the symbol transmitted from co-channel interference sources with unit variance and zero mean, $I_p$ is the number of co-channel interferers, $n_m$ is the AWGN with unit variance and zero mean. The obtained signal at each malicious attacker from the $m^{th}$ antenna is expressed as:

$$R_{m,e} = \sqrt{c_m} h_{m,e} d + \sum_{i_e=1}^{I_e} \sqrt{c_{i_e}} h_{i_e,e} d_{i_e} + n_e \quad (73)$$

where $h_{m,e}$ signifies channel coefficient between the $m^{th}$ antenna and $e^{th}$ malicious observer, $c_{i_e}$ is the intervention power from the $i_e$-th interferer at the intruder, $h_{i_e,e}$ denotes channel coefficient present between the $i_e$-th interferer and the intruder, $d_{i_e}$ shows symbols transmitted from the co-channel interferer with unit variance and zero mean, $I_e$ is the total of all interferers present at the intruder and $n_e$ is the AWGN with unit variance and zero mean at the intruder. The received SINR ($\varrho_{m,e}$, $\varrho_{m,p}$) at the eavesdropper and valid node is given by:

$$\varrho_{m,d} = \frac{\alpha_m |h_{m,d}|^2}{\sum_{i_d=1}^{I_d} \alpha_{i_d} |h_{i_d,d}|^2 + 1} \quad (74)$$

where

$d \in \{p, e\}$, $\alpha_m = \frac{c_m}{N_o}$, $\alpha_{i_d} = \frac{c_{i_d}}{N_o}$ and $c_{i_d} = \eta \, c_m$, $0 \leq \eta \leq 1$

The secrecy capacity denoted by $C_s$ can be given as:

$$C_s = [C_p - C_E]^+ = \max(C_p - C_E, 0) \quad (75)$$

where $C_p = \log(1 + \varrho_p)$ and $C_e = \log(1 + \varrho_e)$, assume CSI of an intruder is not accessible. The BS considers for secure communication, the rate of the channel which under the influence of eavesdropper is given as $(C_E)' = C_p - C_r$ where $C_r$ is the set SR by the BS. The BS attempts to construct the wiretap codes by using $(C_E)'$ and $C_p$. To ensure perfect secrecy rate where secrecy is not compromised $C_r \leq C_s$. An outage of secrecy is said to occur if $C_s < C_r$. This performance of analysis is said to be SOP analysis [126]. The likelihood of SOP under the fading distribution is given by:

$$P_o(C_r) = \Pr[C_s < C_r] \quad (76)$$

or

$$P_o(C_r) = \Pr(C_s < C_r | \varrho_p > \varrho_e) \Pr(\varrho_p > \varrho_e) + \Pr(\varrho_p < \varrho_e) \quad (77)$$

or

$$P_o(C_r) = \int_0^\infty \varrho_p'(2^{C_r}(1+R) - 1) f_{\varrho_e}(R) dR \quad (78)$$

the relationship is said to exist between the CDF and its predefined threshold value ($\varrho_{th}$) as $2^{C_r}(1+R) - 1 \geq \varrho_{th}$ or $2^{C_r}(1+R) - 1 < \varrho_{th}$. The piecewise secrecy outage probability with respect to any bound point $T(\varrho_{th}) = 2^{-C_r}(1+R) - 1$ is given by:

$$P_o(C_r) = \begin{cases} \int_0^{T(\varrho_{th})} \varrho_p'(\varpi_R) f_{\varrho_p}(R) dR \\ \int_{T(\varrho_{th})}^\infty \varrho_p'(\varpi_R) f_{\varrho_p}(R) dR & T(\varrho_{th}) \geq 0 \\ \int_0^\infty \varrho_p'(\varpi_R) f_{\varrho_p}(R) dR & T(\varrho_{th}) < 0 \end{cases} \quad (79)$$

where $\varpi_R = 2^{C_r}(1+R) - 1$. Therefore on the basis of the above-attained closed-form equations, SOP can be calculated, and the factors affecting the SOP include the set secrecy threshold, average SNR of the valid and eavesdropper node, and the number of eavesdropping nodes. From the above mentioned closed-form equations, it can be deduced that interference and the number of interfering devices lead to a decrease in the performance of the security.

Fig. 3. condenses the security attacks in different technologies of 5G along with the perspective of their countermeasures. For D2D communication, the security attacks and preventive aspects of security in inband and outband configurations are briefly mentioned. Further, attacks and the related blocking and precautionary security aspects in the technologies of spectrum sharing, UDN, massive MIMO and IoT are summarized

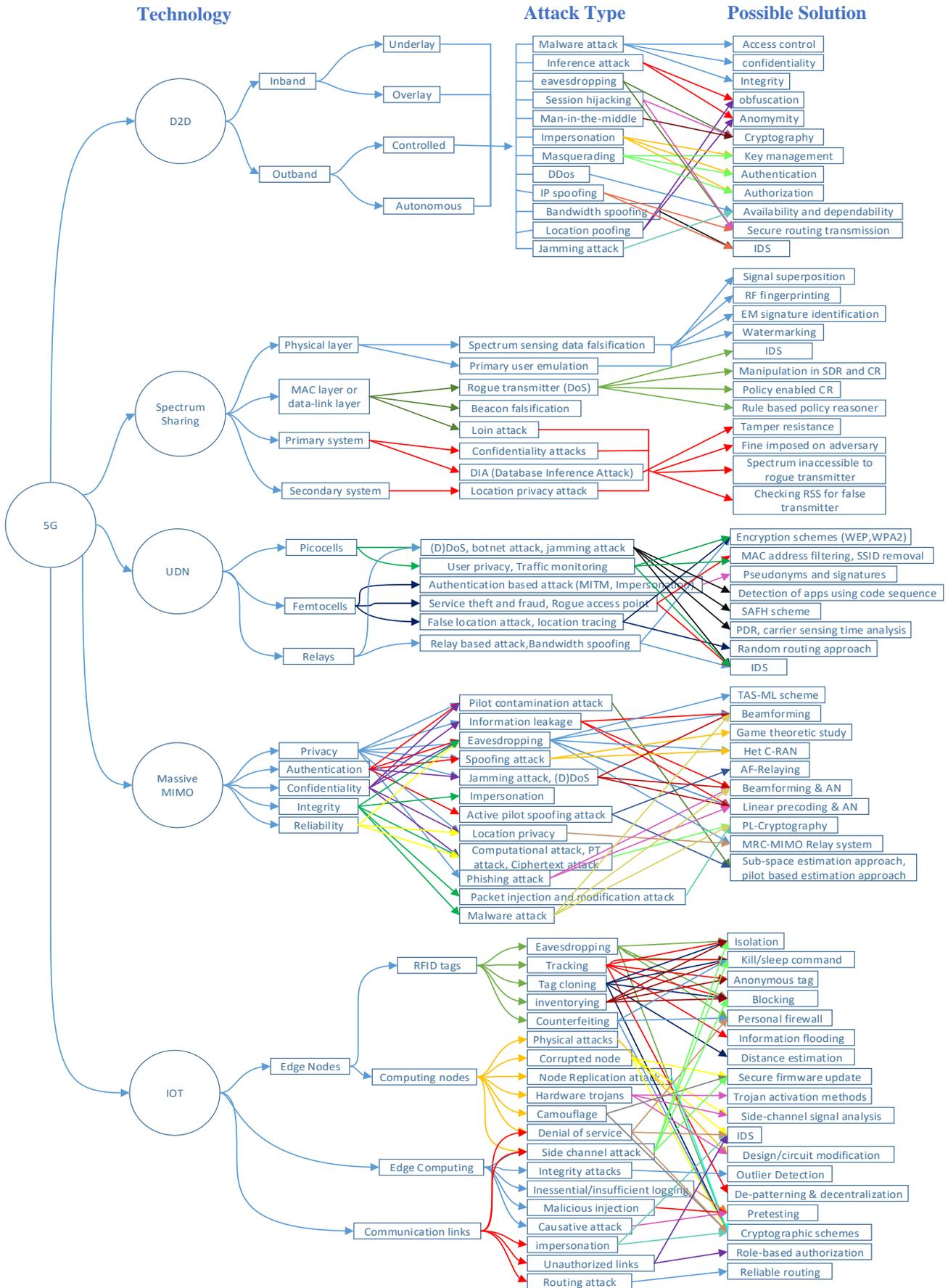

Fig. 3. Visualization of various attacking categories in 5G

## III. THE PROPOSED SCENARIO OF ARTIFICIAL DUST AND ARTIFICIAL RAIN WITH HALF DUPLEX ATTACKING MODEL

The proposed scheme is divided into two main phases. The first phase further comprises of two cases. The first case demonstrates the scenario of AR, and the second case represents the scenario of AD. The second phase involves the modeling D2D half-duplex attack.

### A. Phase First

This phase reveals the impact of attenuation due to AR and AD on WCN and thereby creating a vulnerable environment for the eavesdropper by introducing the attenuation, consequently degrading the communicating channel.

#### 1) Case 1: Scenario of AR (Artificial Rain)

The utmost phenomenon of weather is the rainfall. In the present era of WCN, the attenuation of propagating signals due to rainfall is not equitably approximated. The estimates are reasonably considered for the WCN operating below 6GHz. However, the diameter of the raindrop lies in the range from 0.1mm to 10mm, which is less than the wavelength of the operating signal in WCN. Therefore, creates attenuation in the signal, decreases its strength, and increases path loss, consequently decreases the capacity of the system, which leads to the ultimate decrease in secrecy rate. Thus creates a favorable scenario for the malicious intruder to attack the communicating network. Artificial rain is the phenomenon of causing deliberate rainfall by using the process of precipitation in the form of rain or drizzle or sleet or snow or graupel or hail. This process follows the method of cloud seeding defined as an artificial approach of injecting condensation nuclei typically 0.2 μm or 1/100th the dimensions of a cloud droplet on which water vapor condenses. This mechanism makes use of cloud seeding chemicals such as potassium iodide, silver iodide, liquid propane ammonium nitrate, solid carbon dioxide (dry ice), chloride calcium carbonate, the compound of urea calcium oxide. These chemicals are considered as water-absorbent substances to enhance the formation of the cloud, which increases or induces rainfall. The injection of these chemicals is being dispersed in the clouds through aircraft, dispersion devices positioned on the ground or through remote-controlled rockets, or through generators. There are three methods through which artificial rain can be achieved.

##### a) Static cloud seeding:

This method involves the dispersing of chemical substances such as silver iodide (AgI) into the cloud. This chemical substance exhibits crystalline structure, and due to the difference in vapor pressure, the tiny water droplets or the moisture accumulate on the molecules of the chemical substance, thereby increasing the size of the molecules. Due to sufficient size, the particles enough weighty to fall from clouds and produce the rain.

##### b) Dynamic seeding of a cloud

It includes the precipitation using erect air currents to rise more water droplets to pass through clouds, therefore converting into more rainfall. This method of seeding is considered to be more complex as it involves a sequence of proceedings. It includes 100 times more ice crystals than the static method of cloud seeding.

##### c) Hygroscopic seeding of the cloud

This technique defines the seeding process by scattering salts in the lower part of clouds. The size of salt molecules increases by adjacent combing of water molecules. Due to the increased size and weight, these molecules are expected to precipitate as rainfall. An intruder is examining the communication taking place between the various nodes of defense wireless communication networks and has complete information about the traffic of the communication network. therefore, on the basis of traffic pattern analysis, at any time when intruder finds the essential commands or information that is required to be transmitted to the destined node of the defense communication network, intruder starts the instant procedures of artificial dust, followed by the procedure of artificial rain. Due to the presence of the artificial dust particles or rain particles present in the atmosphere, the interference tends to increase, thereby increases the path loss and thus degrades the communication environment.

#### 2) Case 2: Scenario of AD (Artificial Dust)

AD shows an important aspect that can affect the communicating link. The impact of AD is illustrated by considering the cross-polarization and attenuation on the millimeter and microwave WCNs. The wave attenuation due to dust to AD can be approximated by Rayleigh scattering theory or Mie scattering approximation. AD facilitates an inventive method of communicating with the environment. CARE (Charged Aerosol Release Experiment) represents the AD project handled by NASA to observe the specific concern of artificially generated dust cloud named Noctilucent Clouds. Nevertheless, the potential aspect of AD decreases the secrecy of the network. Therefore, the use of AD by an intruder requires primary attention. The AD is introduced by an

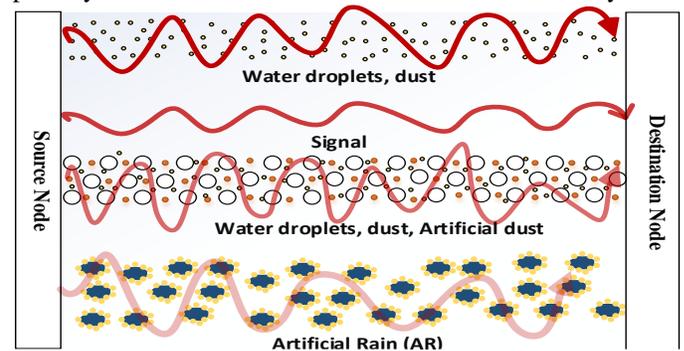
Fig. 4. The impact of artificial dust and artificial rain on the strength of the signal

intruder in the communicating environment of WCN aerially provides the capability to the intruder to examine the communication, sense the information, and thus increases the possibility of eavesdropping in the network. As a result, creates an ability to the intruder to monitor the rate of communication and thus possesses the capability to create hindrances or delay the communication. The impact of AD and AR on the signal strength is shown in Fig. 4. From the source node to the destination node in the absence of AR and AD, the signal strength of the signal occupies minimum losses due to the absence or minimum of obstacles present in the communication channel. Due to the addition of AD, the signal is transmitted through the communication channel by passing through the obstacles creating the decrease of the signal strength, as shown in Fig. 4. However, for AR, the particle size is considerably more than the dust particles, thereby creates a more impact on the signal strength.

From Fig 5 at stage-I, wireless communication environment is devoid of artificial dust or artificial rain, seamless communication is expected to occur. At stage-II, when an intruder introduces the AD particles in the atmosphere, more interference, more path loss is expected to occur. At stage-III, the presence of AR particles creates more interference and attenuation due to the large size molecules of the rain droplets, thereby creating multipath propagation attenuation and thus decreasing the signal strength. The mathematical modeling to determine the impact of AR and AD on WCN can be deduced as:

*a)  Stage I: Free space PL (Path Loss)*

Consider a signal transmitted to a receiver through free space at a distance of 'r'. Suppose there are no hindrances in reference to the receiver and transmitter, and thus the signal is propagating in a straightforward path. The path loss of free space presents a complex factor of scaling in the receiving signal such that the received signal can be calculated as:

$$r(t) = Re \left\{ \frac{\lambda \sqrt{A_l} e^{-\frac{j2\pi r}{\lambda}}}{4\pi r} x'(t) e^{-2\pi f_c t} \right\}, \quad (80)$$

where $A_l$ is the product of field radiation patterns of transmit antenna and receive antenna, $e^{-\frac{j2\pi r}{\lambda}}$ denotes the phase shift that arises due to the traveling distance of wave from transmit antenna to receive antenna 'r', $x'(t)$ denotes the complex baseband signal and is the combination of an in-phase component and quadrature component is given by:

$$x'(t) = u_I(t) + ju_Q(t) \quad (81)$$

where $u_I(t)$ denotes in-phase component and $u_Q(t)$ denotes quadrature component and are represented as:

$$u_I(t) = Re\{x'(t)\} \quad (82)$$
$$u_Q(t) = Im\{x'(t)\} \quad (83)$$

The ratio of the received power to transmitter power is given as:

$$\frac{P_r}{P_t} = \left[\frac{\sqrt{A_l}\lambda}{4\pi r}\right]^2 \quad (84)$$

In the dBm scale

$$P_{r\,(dbm)} = P_{t\,(dbm)} + 10\log_{10}(A_l) + 20\log_{10}(\lambda) - 20\log_{10}(4\pi) - 20\log_{10}(r) \quad (85)$$

Therefore, the free space loss can be expressed as:

$$P_{l\,(db)} = 10\log_{10}\frac{P_t}{P_r} \quad (86)$$

Using equation (84), we get

$$P_{l\,(db)} = -10\log_{10}\frac{A_l \lambda^2}{(4\pi r)^2} \quad (87)$$

Also, the free-space path gain is given by:

$$P_{g\,(db)} = -P_{l\,(db)} \quad (88)$$

$$P_{g\,(db)} = 10\log_{10}\frac{A_l \lambda^2}{(4\pi r)^2} \quad (89)$$

Approximated PL as a function of distance r is expressed as:

$$P_r = P_t\, q \left[\frac{r_o}{r}\right]^{\psi} \quad (90)$$

In dB, the above equation (90) can be rewritten as:

$$P_{r\,(dbm)} = P_{t\,(dbm)} + q_{(db)} - 10\,\psi\log_{10}\left(\frac{r_o}{r}\right) \quad (91)$$

where $q$ is a dimensionless constant depending on the characteristics of average attenuation of channel and antenna, $r_o$ is the distance of reference, $\psi$ is the PL exponent
where $q$ can be expressed as:

$$q_{(db)} = 20\log_{10}\left(\frac{\lambda}{r_o}\right) \quad (92)$$

Using a model of log-normal shadowing, suppose the ratio of transmit power to receive power is represented by '$\phi$' be random, given by log-normal distribution as:

$$p(\phi) = \frac{\Delta}{\sqrt{2\pi}\,\sigma_{\phi\,(db)}\,\phi} \exp\left\{-\frac{(10\log_{10}\phi - \mu_{\phi\,(db)})^2}{2\sigma^2_{\phi\,(db)}}\right\}, \quad \phi > 0 \quad (93)$$

where $\Delta$ is given as:

$$\Delta = \frac{10}{\ln 10} \quad (94)$$

where $\mu_{\phi\,(db)}$ denotes the mean of $\phi_{(db)}$, $\sigma^2_{\phi\,(db)}$ represents the variance of $\phi_{(db)}$ and $\sigma_{\phi\,(db)}$ represents the standard deviation of $\phi_{(db)}$. The expectation of $\phi$ or the linear average path gain or the mean of $\phi$ can be given as:

$$E[\phi] = \mu_\phi = \exp\left[\frac{\mu_{\phi\,(db)}}{\Delta} + \frac{\sigma^2_{\phi\,(db)}}{2\Delta^2}\right] \quad (95)$$

In the dB scale:

$$10\log_{10}\mu_\phi = \mu_{\phi\,(db)} + \frac{\sigma^2_{\phi\,(db)}}{2\Delta^2} \quad (96)$$

Equation (93) can also be written as:

$$p(\phi_{(db)}) = \frac{1}{\sqrt{2\pi}\,\sigma_{\phi\,(db)}} \exp\left\{-\frac{(\phi_{(db)} - \mu_{\phi\,(db)})^2}{2\sigma^2_{\phi\,(db)}}\right\} \quad (97)$$

Therefore the combined path loss and shadowing can be evaluated as:

$$\left(\frac{P_r}{P_t}\right)_{(db)} = 10\log_{10} q + 10\psi\log_{10}\frac{r}{r_o} - \phi_{(db)} \quad (98)$$

In dbm scale,

$$P_{r\,(dbm)} = P_{t\,(dbm)} + q_{(db)} - 10\,\psi\log_{10}\left(\frac{r_o}{r}\right) - \phi_{(dbm)} \quad (99)$$

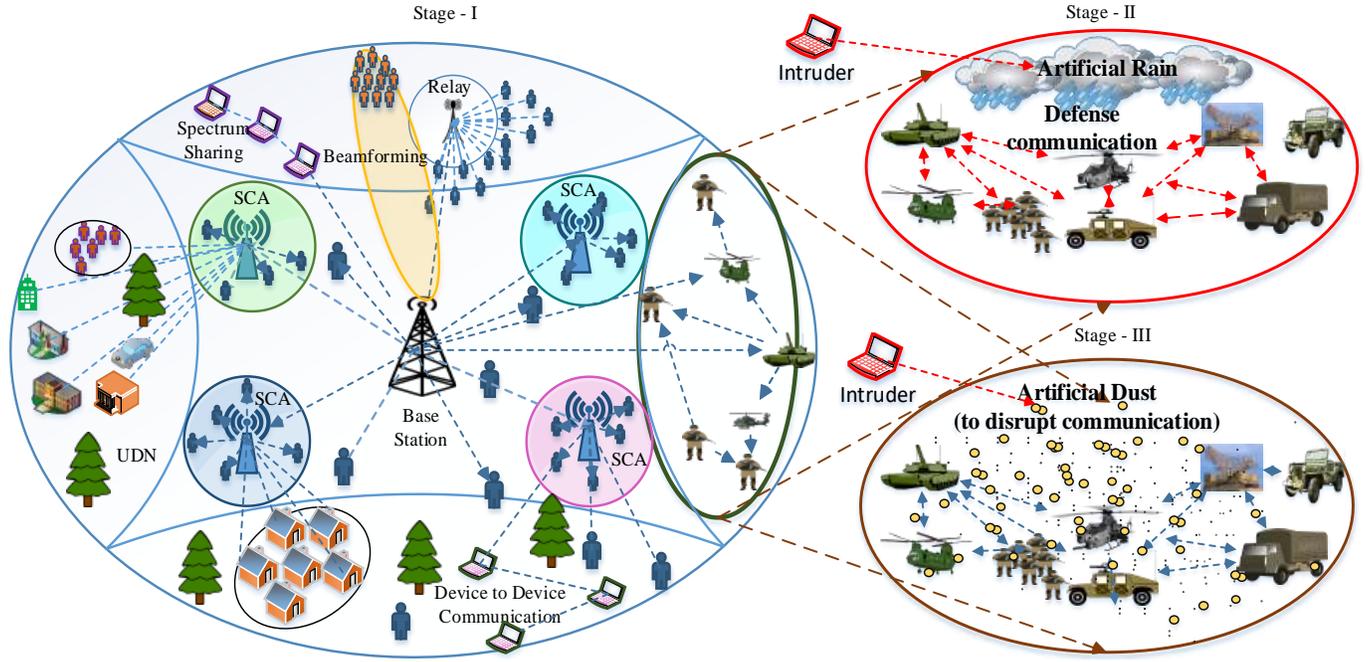

Fig. 5. Impact of artificial rain and artificial dust in the 5G communicating network

*b)    Stage-II: Impact of attenuation on security due to AR*

The attenuation occurring in a signal while traveling through obstacles with depth $\omega$ is approximately given as:

$$j(\omega) = e^{-N\omega} \quad (100)$$

where $N$ denotes the constant of attenuation and depends on the material of the object and its dielectric properties. Equation (100) can also be written as

$$j(\omega_t) = e^{-N\sum_i \omega} \quad (101)$$

$$j(\omega_t) = e^{-N\omega_t} \quad (102)$$

where $\omega_t$ defines the sum of random object depths from which signals travel. Using the central limit theorem, the variable $\omega_t$ can be simplified by approximating the Gaussian random variable. Consider a distance $r_{x,y}$ such that $N_R$ denotes the specific attenuation constant of a signal, due to artificial rainfall.

$$d_R = d_{SC} + d_{AB} + d_{REF} + d_{POL} \quad (103)$$

Such that attenuation due to rainfall can be given as:

$$x'(d_R) = e^{N_R d_R} \quad (104)$$

$$x'(d_R) = e^{N_R d_{SC}} + e^{N_R d_{AB}} + e^{N_R d_{REF}} + e^{N_R d_{POL}} \quad (105)$$

In the dB scale:

$$(x'(d_R))_{db} = N_R d_{SC} + N_R d_{AB} + N_R d_{REF} + N_R d_{POL} \quad (106)$$

where $x'(d_R)$ denotes the attenuation of a signal due to artificial rain caused by an intruder. The equation (100) for path loss, including the effect of artificial attenuation, can be given as:

$$\left(\frac{P_r}{P_t}\right)_{(db)} = 10\log_{10} q + 10\psi \log_{10}\frac{r}{r_o} - \phi_{(db)} - (x'(d_R))_{db} \quad (107)$$

or

$$\left(\frac{P_r}{P_t}\right)_{(db)} = 10\log_{10} q + 10\psi \log_{10}\frac{r}{r_o} - \phi_{(db)} - N_R d_{SC} - N_R d_{AB} - N_R d_{REF} - N_R d_{POL} \quad (108)$$

where, $d_{SC}$ is the scattering depth, $d_{AB}$ is the absorption depth, $d_{REF}$ is the refraction depth, $d_{POL}$ is the polarization depth, $N_R$ specific attenuation that depends on the dielectric properties of the rain droplet and its composition. The specific attenuation in presence of an artificial rainfall can be determined by using power-law relationship given as:

$$N_R = \theta R^\varepsilon \quad (109)$$

where $\theta$ and $\varepsilon$ are the coefficients and depends on the frequency $f$ (in GHz) and lies in ranges from 1 to 6 GHz, 28 – 32GHz and up to 64GHz. Using power-law coefficients (curve-fitting), $\theta$ can also be expressed as:

$$\log_{10}\theta = \sum_{i=1}^{4}\left(\delta_i \exp\left[-\left(\frac{\log_{10} f - \zeta_i}{\vartheta_i}\right)^2\right]\right) + a_k \log_{10} f + b_k \quad (110)$$

where $f$ is the frequency, $a, b, \delta, \zeta,$ and $\vartheta$ are coefficients of $\theta$ $\varepsilon$ can be determined as :

$$\varepsilon = \sum_{i=1}^{4}\left(\delta_i \exp\left[-\left(\frac{\log_{10} f - \zeta_i}{\vartheta_i}\right)^2\right]\right) + a_k \log_{10} f + b_k \quad (111)$$

The coefficient $\theta$ can be determined either for horizontal polarization $\theta_H$ or for vertical polarization $(\theta_V)$. Similarly, $\varepsilon$ can also be determined for horizontal polarization $(\varepsilon_H)$ or vertical polarization $(\varepsilon_V)$. For circular, linear and other path geometries the coefficients $\theta$ and $\varepsilon$ can be calculated as:

$$\theta = \frac{(\theta_H + \theta_V + (\theta_H - \theta_V)\cos^2\alpha \cos 2\beta)}{2} \quad (112)$$

$$\varepsilon = \frac{(\theta_H \varepsilon_H + \theta_V \varepsilon_V + (\theta_H \varepsilon_H - \theta_V \varepsilon_V)\cos^2\alpha \cos 2\beta)}{2\theta} \quad (113)$$

TABLE II
RANGE OF PATH LOSS COMPONENT FOR DIFFERENT SCENARIOS

| | Scenario | Path loss component ($\psi$) |
|---|---|---|
| 1. | Urban macrocells | 3.7 - 6.5 |
| 2. | Building (same floor) | 1.6 – 3.5 |
| 3. | Urban microcells | 2.7 - 3.5 |
| 4. | Building (multiple floors) | 2 - 6 |
| 5. | Home | 3 |
| 6. | Store | 1.8 – 2.2 |
| 7. | Factory | 1.6 - 3.3 |

where $\alpha$ is the angle of path elevation and $\beta$ is the tilt angle of polarization relative to the horizontal

*c)* *Stage-III: Impact of attenuation on security due AD*

From Fig. 5, consider the same distance $r_{x,y}$ from transmitter antenna to receiver antenna such that $V$ denotes the constant of attenuation occurring due to artificial dust $x'(D)$ denotes the attenuation of signal due to artificial dust due to destructive interferences are resulting in a decrease in signal strength. Table II shows the range of path loss components defined for various scenarios, such as macrocells, microcells. From equation (100) the attenuation in a signal while crossing a distance through artificial dust with depth $D$, as the combination of depths occurring due to scattering, absorption and cross-polarization is evaluated as:

$$x'(D) = e^{VD} \qquad (114)$$

The above equation (109) can also be evaluated as:

$$x'(D) = e^{VD_{SC}} + e^{VD_{AB}} + e^{VD_{cp}} \qquad (115)$$

In the dB scale

$$(x'(D))_{db} = VD_{SC} + VD_{AB} + VD_{cp} \qquad (116)$$

Therefore, the attenuation occurring in a signal due to artificial dust can be evaluated for path-loss by using equation (98) as:

$$\left(\frac{P_r}{P_t}\right)_{(db)} = 10\log_{10} q + 10\psi \log_{10}\frac{r}{r_o} - \phi_{(db)} - (x'(D))_{db} \qquad (117)$$

In the dBm scale

$$P_{r(dbm)} = q_{db} + 10\psi \log_{10}\frac{r}{r_o} - \phi_{(db)} - (x'(D))_{dbm} \qquad (118)$$

Such that,

$$(C_s)_i = (C_u - (C_{ev})_i) < C_T \qquad (118a)$$

where $i$ =AR, AD, $C_u$ and $(C_{ev})_i$ is the capacity of the user and intruder (AR, AD), $(C_s)_i$ is the secrecy capacity achieved after AR and AD, less than the threshold capacity $C_T$. Therefore, creating a favorable attacking scenario for the intruder $(C_s)_i$.

*B. Phase Second: Execution of Half-duplex attack in the D2D WCN:*

D2D is one of the emerging technology of wireless communication to satisfy demands as reduced delay, efficient throughput, reduced power consumption. However, D2D wireless communication technology does not involve the direct monitoring of the central network, thus creates a question over the security of D2D communication. Also, D2D technology entails two main stages. The initial stage involves communication through the central network such as BS, and the second stage involves the communication between devices.

D2D communication involves the relay device for communication, consequently gives upsurge to the growth of security issues, as relays are considered to be least secure communication identities in comparison to the BS and SCA. Therefore makes the D2D communication more prone to attacks. In Fig. 6. considering the case of D2D communication with device relaying assisted by the controlled link provided by the BS. A device present at the edge of the cell where the coverage of the signal is weak. Following the D2D communication where another device present in suitable coverage zone of the BS communicates with a device presen at an inadequate coverage, thereby acting as a relay. The BS is responsible for allocating resources to the devices.

The scenario of a half-duplex attack is being modeled by involving three stages. The first stage involves the intrusion of artificial noise to the device present at a poor coverage area by an intruder who continuously examines the communication environment. The intrusion of artificially generated noise creates the effect of jamming while performing D2D communication. The second stage involves the continuous ping at the relay device. This stage involves the mechanism associated with the uplink communication from a relay device to the BS. Authentication of the relay device is being requested during the uplink phase for the allocation of resources. The third stage involves the response received by the intruder from the relay device, followed by the occupancy of the allocated resources. During the downlink communication from the BS to relay device, the intruder pretends as a relay device and grabs the information present in the downlink and takes hold of allocated resources. The mathematical analysis for the HD attack is modeled as:

The equations of state transition can be evaluated by considering the scenario of transmission time intervals for RRC configuration. At time $t_1$, idle RRC_ state of authenticated user equipment makes a request to gNB for RRC setup, as shown in Fig. 7. The setup request consists of an authentication stamp $(k_{t_1})$, identification id $e_{t_1}$, indication bits $C_{t_1}$ specifying proximity indication or Feature Group Indication (FGI). Therefore, for the transmission time interval $t_1$ in the presence of an intruder, the received signal at gNB send by the UE can be defined as:

$$y_{t_1} = (k_{t_1}, e_{t_1}, C_{t_1}, x_{t_1}, h_{t_1})o_{t_1} + n_{t_1} \qquad (119)$$

An assumption is made where an intruder is completely known of channel state information and possesses access to the identity id of the valid user equipment. During the Transition Time Interval (TTI) $t_2$, it is assumed that the intruder is capable of concealing the identity of authenticated user equipment and receives the RRC setup in the downlink phase. The received signal by the intruder instead of valid UE can be expressed as:

$$y_{t_{2ev}} = (k_{t_2}, e_{t_2}, C_{t_2}, x_{t_2}, a_{t_2}, h_{t_{2ev}})o_{t_2} + n_{t_{2ev}} \qquad (120)$$

At the same transition time interval, intruder transmits the RRC setup signal along with a random sequence to the authenticated UE. The received signal by the valid UE sent by an intruder is given by:

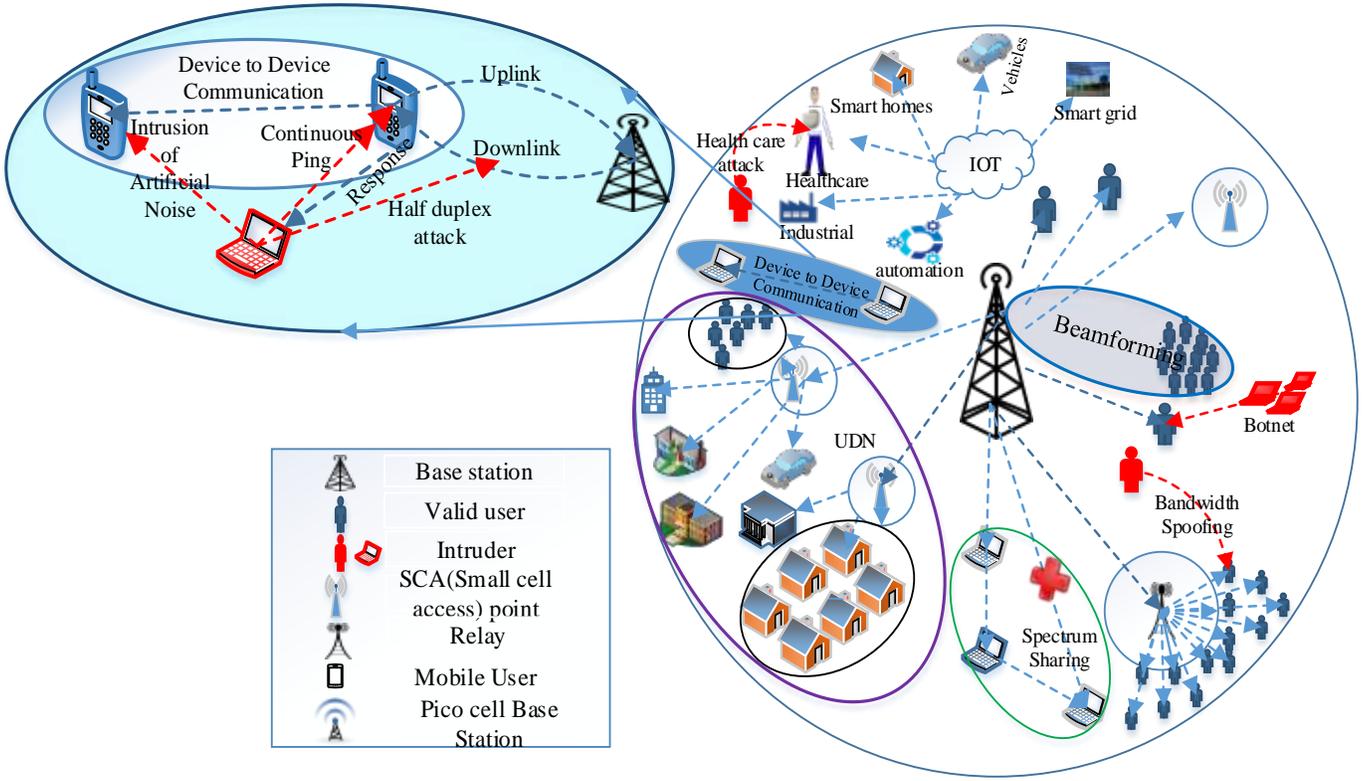

Fig. 6. A scenario of Half-Duplex (HD) attack

$$y_{t_2} = (k_{t_2}, e_{t_2}, C_{t_2}, x_{t_2}, a_{t_2}, h_{t_2}, r_{t_2})o_{t_2} + n_{t_2} \quad (121)$$

For the case of a connection failure mechanism, such that intruder again occupies the idle state and repeats the process. The attacking mechanism of Half- Duplex attack is defined as the involvement of eavesdropping of sole downlink process for the TTI $t_2$ instead of attacking both the phases, i.e., UL and DL. The probability of occurrence of Half-Duplex attack can be evaluated as:

Considering two possible scenarios for downlink reception:
1. Successful reception of DL RRC setup by the valid UE denoted $p_{DL}$ (success)
2. Successful reception of DL RRC setup by an intruder denoted by $1 - p_{DL}$ (failure)

$$y_{t_n}[t_n] = y_{t_1}[t_1], y_{t_2}[t_2], y_{t_3}[t_3], \dots, y_{t_N}[t_N] \quad (122)$$

where n=1,2,3, …, N. N denotes Nth natural number. $y_{t_n}[t_n]$ is considered to be a discrete-time process whose density function is given by:

$$f_{y_{t_n}[t_n]}(y_{t_n}[t_n]) = q_{DL}\delta(y_{t_n}[t_n]) + p_{DL}\delta(y_{t_n}[t_n] - 1) \quad (123)$$

where $\delta(.)$ depicts unit impulse function, $q_{DL} = 1 - p_{DL}$ represents the condition of failure. The second order density function for the received signal $y_{t_n}[t_n]$ can be expressed as:

$$f_{y_{t_1}[t_1]y_{t_2}[t_2]}(y_{t_1}[t_1]\,y_{t_2}[t_2]) = q_{DL}{}^2\delta(y_{t_1}[t_1])\delta(y_{t_2}[t_2]) + p_{DL}q_{DL}\delta(y_{t_1}[t_1] - 1)\delta(y_{t_2}[t_2] - 1) \quad (124)$$

At the authenticated UE, for $t_n$ time intervals, the sum of the received signal is given by:

$$S(y_{t_n}[t_n] = u) = y_{t_1}[t_1] + y_{t_2}[t_2] + y_{t_3}[t_3] + \cdots + y_{t_N}[t_N] \quad (125)$$

Therefore, the probability of successful reception at the valid UE for TTI $t_n$, can be expressed as:

$$P_{DL}(S(y_{t_n}[t_n] = u)) = \binom{n}{u} p_{DL}^u\, q_{DL}^{n-u} \quad (126)$$

Similarly, for TTI $t_n$ the probability of successful reception at an intruder or the probability of HD attack can be expressed as:

$$P_{DLev}(S(y_{t_n}[t_n] = u)) = \binom{n}{u} q_{DL}^u\, p_{DL}^{n-u} \quad (127)$$

For the scenario, when the intruder attacks both UL and DL during RRC state transition from idle to connected. For UL RRC request from UE to gNB at TTI $t_1$, from equation (119) the received signal by an intruder can be expressed as:

$$y_{t_1 ev} = (k_{t_1}, e_{t_1}, C_{t_1}, x_{t_1}, h_{t_{ev}})o_{t_1} + n_{t_{ev}} \quad (128)$$

From equation (125) and (126), equation (128) can also be represented for the condition of successful transmission to gNB ($p_{UL}$) as:

$$P_{UL}(S(y_{t_n}[t_n] = u)) = \binom{n}{u} p_{UL}^u\, q_{UL}^{n-u} \quad (129)$$

where $q_{UL} = 1 - p_{UL}$, for the scenario where the successful transmission to intruder takes place, is expressed as:

$$P_{ULev}(S(y_{t_n}[t_n] = u)) = \binom{n}{u} q_{UL}^u\, p_{UL}^{n-u} \quad (130)$$

$$P_{total} = P_{UL}(S(y_{t_n}[t_n])) + P_{DL}(S(y_{t_n}[t_n])) - P_{UL}(S(y_{t_n}[t_n])) \cap P_{DL}(S(y_{t_n}[t_n])) \quad (131)$$

Since, $y_{t_n}[t_n]$ is IID., therefore,

$$P_{UL}(S(y_{t_n}[t_n])) \cap P_{DL}(S(y_{t_n}[t_n])) = 0 \quad (132)$$

Thus, the above equation can be expressed as:

TABLE III
COMPARISON OF THE PROPOSED SCHEME WITH EXISTING SCHEMES

| Ref. | Objective | Associate parameters | Attacking scenario | | | | Dust | | Rain | |
|---|---|---|---|---|---|---|---|---|---|---|
| | | | Analysis | | Security impact | Half Duplex attack | General Analysis | Security impact | General Analysis | Security impact |
| | | | DL | UL | | | | | | |
| [92] | Single and multi-channel downlink resource management for the improvement of physical layer security | D2D sum rate, security probability | ● | ○ | ● | ○ | ○ | ○ | ○ | ○ |
| [151] | Enhancement of the security by utilizing the interference caused by D2D to cellular users | D2D sum rate, probability of CU security QoS is satisfied | ○ | ● | ● | ○ | ○ | ○ | ○ | ○ |
| [152] | Security enhancement by analyzing the inference created by D2D communication using accurate closed-form bounds | Secrecy probability of cellular link, SINR, connection probability of cellular link | ● | ○ | ● | ○ | ○ | ○ | ○ | ○ |
| [153] | Impact of rain on the propagation of millimeter waves at 26 GHz | Rain attenuation, specific attenuation | ○ | ○ | ○ | ○ | ○ | ○ | ● | ○ |
| [154] | Effect of atmospheric gases, rain, foliage, and diffraction at 28GHz, 30GHz and 60GHz | Specific attenuation, fog attenuation, foliage attenuation, rain attenuation | ○ | ○ | ○ | ○ | ○ | ○ | ○ | ○ |
| [155] | Statistics of rain attenuation over terrestrial paths for the cell size of 200m | Rain attenuation, excedance probability, predicted FMT gain | ○ | ○ | ○ | ○ | ○ | ○ | ● | ○ |
| [156] | Impact of rain on achievable rates and capacities of millimeter wave MIMO system | Rain attenuation, capacity, bit rate | ○ | ○ | ○ | ○ | ○ | ○ | ● | ○ |
| [157] | Study of phase delay and attenuation caused by dust and sand storms using Ghobrial et al. 's formula | Attenuation, visibility | ○ | ○ | ○ | ○ | ● | ○ | ○ | ○ |
| [158] | Effect of dust and sand storms on millimeter wave signal attenuation | Specific attenuation, propagation factor | ○ | ○ | ○ | ○ | ● | ○ | ○ | ○ |
| [159] | Calculation of wave attenuation in the storms of rain and dust at a frequency of 10-100GHz | Averaged extinction cross-section (ECS), attenuation | ○ | ○ | ○ | ○ | ● | ○ | ○ | ○ |
| [160] | Effectt of dust and sand storms on propagation of microwaves | Attenuation, cross polar discrimination | ○ | ○ | ○ | ○ | ● | ○ | ○ | ○ |
| [161] | Effect of dust on FSO system of communication in semi-arid and arid atmosphere | Attenuation, specific attenuation, visibility | ○ | ○ | ○ | ○ | ● | ○ | ○ | ○ |
| [162] | An equation for radar based on parabolic wave equation is computed in the presence of dust and sand storm for the prediction of backscattering millimeter wave | Power ratios | ○ | ○ | ○ | ○ | ● | ○ | ○ | ○ |
| [163] | To observe the effect of dust and sand-dust storms in earth satellite links on microwave propagation | Cross polarization discrimination, attenuation, attenuation coefficient | ○ | ○ | ○ | ○ | ● | ○ | ○ | ○ |
| proposed | To observe the impact of rain, dust and half duplex attack at a frequency 10-100 GHz | Secrecy rate, secrecy non-outage probability, capacity, distance, attenuation, specific attenuation | ● | ● | ● | ● | ● | ● | ● | ● |

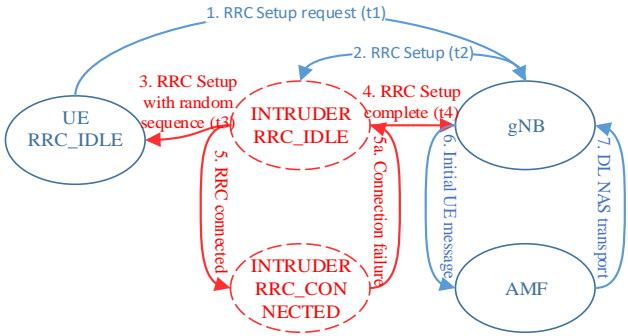

Fig. 7. Half-Duplex attack using RRC setup

$$P_{total} = P_{UL}(S(y_{t_n}[t_n])) + P_{DL}(S(y_{t_n}[t_n])) \quad (133)$$

Now, the probability of miss-rate can be obtained as:

$$P(miss-rate_{FD}) = \frac{(1-P_{ULev})+(1-P_{DLev})}{(P_{Total})} \quad (134)$$

where FD denotes the Full-Duplex mode, i.e., intruder trying to eavesdrop both during uplink and downlink.
Similarly,

$$P(mis-rate_{HD}) = \frac{(1-P_{DLev})}{(P_{Total})} \quad (135)$$

Comparing equation (134)and (135), we get:

Such that $P(mis-rate_{FD}) > P(mis-rate_{HD})$

From equation (135), it can be concluded that the miss-rate of the Full Duplex attack is greater than the miss-rate of HD attack. Thereby increasing the chance of the intruder to attack the network by the execution of HD attack as compared to the full-duplex attacking mechanism.

Table III provides a brief comparison of related study and the proposed methodology in view of security. It defines the objective of previously attempted work along with the considered associated parameters of security.

## IV. CHALLENGES IN CONTEXT OF SECURITY

Regardless of the existing work on the security of 5G WCN, various significant research challenges are required to be addressed to prevent compromised features of the security. This section describes some present and future research directions in view of security of WCN

### A. Secrecy capacity (SC) and SOP

SC and SOP are the two essential performance parameters of security. Based on the security impact of the network, these two parameters are analyzed. SC is precisely defined as the difference between the capacity of the main channel and capacity of the channel malicious intruder. It defines the maximum rate at which data can be communicated from the sender to the receiver under the appropriate performance of secrecy. Security capacity is major affected by the presence of the eavesdropper. As the number of eavesdroppers increase, secrecy capacity inclines to decrease. In the case of 5G WCN, SC and SOP are examining parameters that primarily affect the performance of the network, including a D2D communication, UDN. However, these two factors also affect the performance of other technologies of 5G WCN. SOP defines the probability of an outage when the SC is smaller than a certain defined value. SOP tends to progress with the rise in the number of malicious intruders.

### B. Resource spoofing

Due to increased vulnerable breaches in 5G communication, network attacks are prominently increasing day by day. These breaches led to catching sight for the intruder to attack. The interest of an attacker can be either of active type or of the passive type. The resource spoofing is the fundamental attention grabbing the interest of an attacker to attack the user and spoof the resources allocated to the legitimate user. It targets control channels to spoof allocated resources. It primarily affects UDN, D2D technologies of 5G. Therefore, security performance is required to be enhanced by proper mechanisms to prevent the inefficient, illegal, and unproductive usage of resources by an attacker.

### C. Complexity

As the demands of the users increase nowadays, which correspondingly increases the deployment of the increased number of access points, the number of cells. This leads to the development of UDNs, Massive MIMO technologies of 5G WCN. These deployed structures require proper management and control for their appropriate functioning to fulfill the seamless connectivity of links. This results in an upsurge in a number of divisions present in the WCN, creating suitable breaches for a malicious user to attack. Therefore, creates a security issue in existing technologies of 5G WCN.

### D. Identity spoofing

One of the chief challenges in the current security of the 5G WCN is the attack on the identity of the authentic users by malicious attackers. In the case of UDN, D2D identity spoofing is one of the chief challenges which is required to be encountered. As an identity parameter is included in the primary aspect of the security, such as authentication, authorization of authorized users. Therefore, the security of identity is required to enable valid users to authorized access to the authenticated resources at the right time while maintaining integrity.

### E. CSI availability

The availability of CSI is another challenge that is required to determine the security performance of the network. In case UDN, Massive MIMO, D2D, availability of CSI of users necessitates the secure transmission of the information. Security tends to increase with an increase in the accuracy of CSI.

### F. Handover management attacks

In 5G WCN, UDN consists of the bulky number of microcells, picocells, femtocells to fulfill the growing demands of the subscribers. Consequently, with an increased number of cells, handover tends to increase frequently. As handover site is the most appropriate spot where an attack is most probable to happen due to a decrease in secrecy rate. Therefore, in technologies of 5G such as UDN, Massive MIMO suitable handover management is required to prevent the sensitive site of handover from being attacked and thus increasing the security of the network.

*G. Latency*

One of the chief issues that deteriorate the functionality of the 5G security network is latency. It creates a key impact on the service of security to the fundamental parameters of the security, including integrity, confidentiality, and authentication. The users existing in the 5G network necessitate high pace connectivity with uninterrupted communication links. However, due to latency, the delay is created in the communication network followed by its consequential effect with an adequate increase of chance for the intruders to examine or eavesdrop the communication between the transmitter and receiver. By reason of its effect, the threat is introduced to the security environment of the communication network.

*H. Jamming attack*

An important parameter that affects the performance of the security of the 5G network. Jamming in UDN, D2D gives an optimal decrease in the security of these technologies of 5G. In a jamming attack, an attacker sends more powered signals using an equivalent frequency range, thereby disrupting legitimate communication between transmitter and receiver.

These challenges are the chief factors that affect the security of the 5G network. Therefore, the solution to these problems provides a challenging aspect in the field of research and academia to provide secure communication. Several solutions were approached to countermeasure the challenges mentioned above to enhance the performance of the security. For increased complexity of the network, beamforming tactic has been adopted. Various mechanisms of intrusion detection are proposed such as Adaptive Intrusion Detection system (AIDS) using HMM (Hidden Markov Model) for bandwidth spoofing, Intrusion Detection for using Adaptive Neural Fuzzy Inference System, handover management schemes such as TPA, RBA for the security and protection of most feasible attacking zone of the network.

## V. CONCLUSION

This paper unifies the comprehensive survey based on a security perspective in different technologies of 5G NR, followed by the investigation of their security solutions. The generalized security model for the technologies of 5G NR is expressed in a systemized manner. Moreover, the integration of these technologies raises additional security issues, specifically in terms of authentication, confidentiality, privacy, and integrity. Further, a schematized attacking scenario is proposed by operating the concept of artificial rain or artificial dust, for which the security model is derived on account of the secrecy rate parameter. This implicates the scenario for HD attack where an intruder, instead of targeting both uplink and downlink, attacks the downlink exclusively to obtain the allocated resources that were meant to be allocated to the authenticated user. Therefore, adaptive and novel mechanisms of security are required to be acquired for the development of security services in 5G wireless communication networks

APPENDIX
TABLE IV
LIST OF ACRONYMS

| S.no. | Acronym | Full form |
|---|---|---|
| 1. | AD | Artificial Dust |
| 2. | ANFIS | Adaptive Neuro-Fuzzy Inference System |
| 3. | AR | Artificial Rain |
| 4. | AP | Access Point |
| 5. | AN | Artificial Noise |
| 6. | AWGN | Additive White Gaussian Noise |
| 7. | BB | Beam Broadening |
| 8. | BM | Beam Merging |
| 9. | CSI | Channel State Information |
| 10. | CBRS | Citizens Broadband Radio Service |
| 11. | D2D | Device to Device |
| 12. | DoS | Denial of Service |
| 13. | ECS | Extinction Cross section |
| 14. | FMT | Fade Mitigation Techniques |
| 15. | FGI | Feature Group Indication |
| 16. | GSC | General Selection Combining |
| 17. | IDS | Intrusion Detection System |
| 18. | IoT | Internet of Things |
| 19. | MCC | Minutia Cylinder Code |
| 20. | MDL | Minimum Description Length |
| 21. | MIMO | Multiple Input Multiple Output |
| 22. | ML | Machine Learning |
| 23. | MMSE | Minimal Mean Square Error |
| 24. | MRC | Maximal Ratio Combining |
| 25. | MSRS | Minimal Self-interference Relay Selection |
| 26. | NN | Neural Networks |
| 27. | NOMA | Non Orthogonal Multiple Access |
| 28. | ORS | Optimal Relay Selection |
| 29. | OFDMA | Orthogonal Frequency Division Multiple Access |
| 30. | PDF | Probability Density Function |
| 31. | PET | Privacy Enhancing Technology |
| 32. | PLA | Physical Layer Authentication |
| 33. | PLS | Physical Layer Security |
| 34. | PRS | Physical Relay Selection |
| 35. | PSM | Password Strength Meters |
| 36. | PUE | Primary User Emulation |
| 37. | PUF | Physical Unclonable Functions |
| 38. | QoS | Quality of Service |
| 39. | RBA | Region Based Algorithm |
| 40. | RRC | Radio Resource Control |
| 41. | RTC | Request To Cooperate |
| 42. | RSC | Random Selection Combining |
| 43. | SAFH | Secret Adaptive Frequency Hopping |
| 44. | SCA | Small Cell Access |
| 45. | SINR | Signal to Interference plus Noise Ratio |
| 46. | SLS | Sector Level Sweep |
| 47. | SNR | Signal to Noise Ratio |
| 48. | SOP | Secrecy Outage Probability |
| 49. | SR | Secrecy Rate |
| 50. | SSID | Service Set Identifier |
| 51. | SVM | Support Vector Machines |
| 52. | SWIPT | Simultaneous Wireless Information and Power Transfer |
| 53. | TAS | Transmit Antenna Selection |
| 54. | TPA | Traffic Pattern Analysis |
| 55. | TSD | Threshold based Switched Diversity |
| 56. | TTI | Transition Time Interval |
| 57. | UDN | Ultra Dense Network |
| 58. | UE | User Equipment |
| 59. | WCN | Wireless Communication Network |
| 60. | ZF | Zero Forcing |

TABLE V
CURRENT SECURITY PROJECTS IN WCN

| S.No. | Project name | Aim of research | Area of research | HTTP location |
|---|---|---|---|---|
| 1. | NetGuard Security Management | To secure and protect physical, virtual, and hybrid mobile networks. | Network integrity | https://networks.nokia.com/solutions/security-management |
| 2. | 5G ensure | To achieve trustworthy, viable and secure 5G network | Authentication, privacy, trust, security monitoring, network management, and virtualization | http://www.5gensure.eu/achievements |
| 3. | ANASTACIA(Advanced Networked Agents for Security and Trust Assessment in CPS (Cyber-Physical Systems) / IOT Architectures) | to identify cyber-security pertaining to researching, developing and demonstrating a holistic solution capable of trust and security by design for CPS based on IoT and Cloud architectures. | Security Orchestration intelligently Leveraging and SDN and NFV | http://www.anastacia-h2020.eu/ |
| 4. | Nokia Threat Intelligence Center | To detect and identify malware infections | Detection and identification of malware on the basis of detection rules, command-and-control communication, and other network behavior | https://networks.nokia.com/solutions/threat-intelligence |
| 5. | ECLEXYS Cybersecurity | To oppose attacks and analyze intrusion detection | Data security by modular firewall and intrusion detection system | https://eclexys.com/cybersecurity/ |
| 6. | Cybersecurity | Locate vulnerabilities and access points into an organization's cyber networks, web application server, WIFI networks | Protection of the cyberinfrastructure | https://spectorsecurity.com/services/cyber-security/ |
| 7. | Radware | To protect against the advanced threats | Protection mechanism using machine learning based on cybersecurity | https://www.radware.com/ |
| 8. | CyberSec4Europe | To address cyber threats and security problems | Cybersecurity competence network | https://www.nics.uma.es/projects/cybersec4europe |
| 9. | SMOG- Security Mechanisms for fog cOmputinG | To provide trusted interaction between the entities present in a fog ecosystem | Fog security | https://www.nics.uma.es/projects/smog |
| 10. | DISS-IIoT Design and Implementation of Security Services for the Industrial Internet of Things | To create innovative security services to determine probable security gaps and cover illegal access | Internet of things and critical infrastructure | https://www.nics.uma.es/projects/diss-iiot |

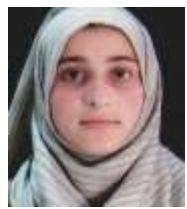

**Misbah Shafi** received the B.E degree in electronics and communication engineering from Islamic University of science and technology, Jammu and Kashmir, India, in 2016 and M.Tech degree in electronics and communication from Lovely Professional University, Punjab, India, in 2018. She is currently pursuing a Ph.D. degree in electronics and communication engineering at Shri Mata Vaishno Devi University, Jammu and Kashmir. Her research interest includes network security in emerging technologies of 5G wireless communication networks, cryptography, optical fibre communication networks. Currently she is doing her research on security issues of 5G NR. She is working on MATLAB for wireless communication.

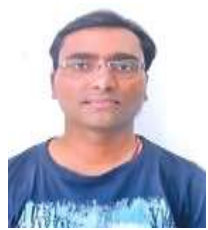

**Dr. Rakesh K Jha (S'10, M'13, SM 2015)** is currently an Associate Professor in the School of Electronics and Communication Engineering, Shri Mata Vaishno Devi University, Katra, Jammu and Kashmir, India. He is carrying out his research in wireless communication, power optimizations, wireless security issues, and optical communications. He has done B.Tech in Electronics and Communication Engineering from Bhopal, India and M. Tech from NIT Jalandhar, India. He received his Ph.D. degree from NIT Surat, India, in 2013. He has published more than 41 Science Citation Index Journals Papers, including many IEEE Transactions, IEEE Journal, and more than 25 International Conference papers. His area of interest is Wireless communication, Optical Fiber Communication, Computer networks, and Security issues. Dr. Jha's one concept related to router of Wireless Communication was accepted by ITU (International Telecommunication Union) in 2010. He has received young scientist author award by ITU in Dec 2010. He has received an APAN fellowship in 2011, 2012, 2017, and 2018 and student travel grant from COMSNET 2012. He is a senior member of IEEE, GISFI and SIAM, International Association of Engineers (IAENG), and ACCS (Advanced Computing and Communication Society). He is also member of ACM and CSI, many patents, and more than 1765 Citations in his credit.

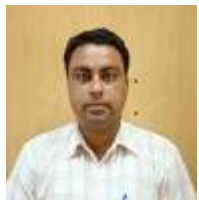

**Dr. Manish Sabraj** is currently an Assistant Professor in the School of Electronics and Communication Engineering, Shri Shri Mata Vaishno Devi University, Katra, Jammu and Kashmir, India. He has received his B.Tech degree in Electronics and Communication from Govt. College of Engineering and Technology Jammu and Kashmir, India, in 2001 and M.Tech degree in Electronics and Communication from IIT Guwahati, Guwahati, India, in 2003. He has completed his Ph.D. degree from Shri Mata Vaishno Devi University, Jammu and Kashmir, India in 2013. His research interest includes Digital Signal Processing, Wireless Sensor Networks, and Spectrum estimation and analysis.